\documentclass[fleqn,usenatbib]{mnras}

\usepackage{newtxtext,newtxmath}

\usepackage[T1]{fontenc}

\DeclareRobustCommand{\VAN}[3]{#2}
\let\VANthebibliography\thebibliography
\def\thebibliography{\DeclareRobustCommand{\VAN}[3]{##3}\VANthebibliography}

\usepackage{graphicx}	
\usepackage{amsmath}	
\usepackage{gensymb}	
\usepackage{xcolor}

\newcommand{\thisstar}{HD~190412}
\newcommand{\thisstarA}{HD~190412~A}
\newcommand{\thisstarAa}{HD~190412~Aa}
\newcommand{\thisstarAB}{HD~190412~AB}
\newcommand{\thisstarC}{HD~190412~C}

\newcommand\masyr{\ensuremath{\text{mas}\,\text{yr}^{-1}}}
\newcommand\Ne{\ensuremath{^{22}\text{Ne}}}

\definecolor{my_color}{HTML}{CF0000}                            
\newcommand\boldnew{\textcolor{black}}
\newcommand\boldagain{\textcolor{black}}

\defcitealias{Fuhrmann19}{F19}
\defcitealias{Tokovinin20}{T20}
\defcitealias{AlmeidaFernandes18}{A18}
\defcitealias{Veyette18}{V18}
\defcitealias{GentileFusillo19}{G.F. et al. (2019)}
\defcitealias{GentileFusillo21}{G.F. et al. (2021)}

\title[Crystallizing White Dwarf in Sirius-like System]{A Crystallizing White Dwarf in a Sirius-Like Quadruple System}

\author[Venner et al.]{Alexander Venner,$^{1}$\thanks{E-mail: alexandervenner@gmail.com}
Simon Blouin,$^{2}$
Antoine Bédard,$^{3,4}$
Andrew Vanderburg$^{5}$
\\
$^{1}$Centre for Astrophysics, University of Southern Queensland, Toowoomba, QLD 4350, Australia\\
$^{2}$Department of Physics and Astronomy, University of Victoria, Victoria, BC V8W 2Y2, Canada\\
$^{3}$Department of Physics, University of Warwick, Coventry, CV4 7AL, UK\\
$^{4}$Département de Physique, Université de Montréal, Montréal, QC H3C 3J7, Canada\\
$^{5}$Department of Physics and Kavli Institute for Astrophysics and Space Research, Massachusetts Institute of Technology, Cambridge, MA 02139, USA\\
}

\date{Accepted 2023 June 02. Received 2023 June 02; in original form 2022 September 21}

\pubyear{2023}

\begin{document}
\label{firstpage}
\pagerange{\pageref{firstpage}--\pageref{lastpage}}
\maketitle

\begin{abstract}

The observational signature of core crystallization of white dwarfs has recently been discovered. However, the magnitude of the crystallization-powered cooling delay required to match observed white dwarfs is larger than predicted by conventional models, requiring additional mechanisms of energy release in white dwarf interiors. The most ideal benchmarks for understanding this discrepancy would be bright and nearby crystallizing white dwarfs with \boldagain{total ages that} can be externally constrained. In this work we report that a recently discovered white dwarf is a bound companion to the triple star HD~190412, forming a new Sirius-like system in the solar neighbourhood. The location of HD~190412~C on the $T_{\text{eff}}-\text{mass}$ diagram implies it is undergoing crystallization, making this the first \boldagain{confirmed} crystallizing white dwarf whose \boldagain{total} age can be externally constrained. Motivated by the possibility that a cooling delay caused by crystallization can be directly detected for this white dwarf we employ a variety of methods to constrain the age of the system; however, our empirical age anomaly of $+3.1\pm1.9$~Gyr is ultimately too imprecise to reach statistical significance, preventing us from making strong constraints to models of white dwarf crystallization. Our results are nonetheless compatible with the recent hypothesis that $^{22}$Ne phase separation is responsible for the excess cooling delay of crystallizing white dwarfs. The discovery of this system at only 32~parsecs suggests that similar benchmark systems are likely to be common; future discoveries may therefore provide powerful tests for models of white dwarf crystallization.

\end{abstract}




\section{Introduction} \label{sec:intro}

White dwarfs are the degenerate remnants of stars with initial masses below $\lesssim$ 8 $M_\odot$ that have shed their outer layers following the end of stellar nucleosynthesis. Typical white dwarfs are composed of an ionised C/O core surrounded by thin exterior layers of H and He. While the core is initially liquid, as the white dwarf ages and cools the core will eventually undergo crystallisation and transform into a solid state. \citet{vanHorn68} was the first to postulate that this phase transition would release a significant amount of latent heat, counteracting the otherwise relatively monotonous cooling of the white dwarf, and therefore predicted the existence of distinct sequences of crystallising white dwarfs on the Hertzsprung-Russell (H-R) diagram. However, the paucity of precise luminosity measurements for white dwarfs prevented the direct detection of this phenomenon for several decades.

\textit{Gaia} is an ongoing space mission designed to measure precise astrometry for a billion stars across the sky \citep{Gaia}. Among many other science results, the second \textit{Gaia} data release \citep[\textit{Gaia} DR2,][]{GaiaDR2} provided parallaxes for tens of thousands of white dwarfs, greatly increasing the number of measured distances and luminosities for white dwarf stars. \citet{Tremblay19} used this data to identify a sequence of white dwarfs lying transverse to the typical cooling sequence, spanning across the bulk of the colour-magnitude diagram for local white dwarfs, and concluded that this feature can only be explained by core crystallisation. Crystallisation produces a pile-up of white dwarfs with reduced cooling rates whose position on the H-R diagram is mass-dependent, with more massive white dwarfs crystallising at higher temperatures due to their higher core densities. \citet{Tremblay19} recognised that in addition to the latent heat released by crystallisation, phase separation in the C/O core contributes significantly to the cooling delay because the sedimentation of O forces fluid C to rise upwards, increasing the rate of energy release during crystallisation. However, even accounting for both of these effects their models do not completely reproduce the luminosity distribution of crystallising white dwarfs, with the observed pile-up being both narrower and more significant than predicted. This indicates that white dwarf cooling models underestimate the delay in cooling caused by crystallisation, implying that there are additional mechanisms for energy release in the interiors of crystallising white dwarfs that are not accounted for by traditional models \citep{Blouin20}.

\citet{Cheng19} identified a more severe discrepancy of the same type at the massive end of the crystallisation sequence, which they name the Q-branch after \citet{GaiaDR2HRD}. The authors found that a remarkably large cooling delay of $\sim$8~Gyr must apply to $\sim$6\% of high-mass ($1.08-1.23~M_\odot$) white dwarfs to explain the observed population of Q-branch white dwarfs. \citet{Cheng19} suggest that a plausible cause for the additional cooling delay is \Ne{}. \Ne{} is the most significant chemical impurity in the cores of C/O white dwarfs, with an expected mass fraction of X(\Ne{})~$=0.014$ for descendants of solar-metallicity progenitor stars \citep{Isern97, Cheng19}, and can affect energy transport in the interiors of crystallising white dwarfs in two ways. The first is sedimentation (or gravitational settling) in the liquid phase \citep{Bildsten01, Deloye02}, a mechanism which has previously been invoked to explain the white dwarf cooling delay observed in the old open cluster NGC~6791 \citep{GarciaBerro10, Althaus10}. While \Ne{} sedimentation can produce a significant cooling delay, it is difficult to reproduce the great magnitude of the delay observed in Q-branch white dwarfs. \citet{Camisassa21} find that \Ne{} mass fractions enhanced as much as X(\Ne{}) = 0.06 can reproduce the large cooling delay, but it is not evident that such high abundances of \Ne{} are possible for Q-branch white dwarfs. \citet{Bauer20} suggest that separation of \Ne{} into solid clusters which then undergo rapid gravitational settling can explain the cooling delay, however \citet{Caplan20} subsequently demonstrated that it is not possible for these clusters to form within C/O white dwarfs.

The second mechanism by which \Ne{} can produce a significant cooling delay is through phase separation \citep{Isern91, Segretain96, Blouin21}. \Ne{} phase separation can generate a cooling delay because solid crystals in the core of a crystallising white dwarf may be incidentally depleted in \Ne{} compared to the surrounding liquid; if this depletion is great enough this can cause the crystals to rise upwards, simultaneously liberating gravitational energy and displacing \Ne{}-rich liquid towards the core, which acts to slow crystallisation of the interior \citep{Blouin21}. This distillation process results in \Ne{} being preferentially stratified at lower layers until the free \Ne{} supply in the core is exhausted, at which point \Ne{} phase separation stops and the remaining (\Ne{}-depleted) C/O liquid will continue crystallisation as usual. \citet{Blouin21} found that the effects of \Ne{} phase separation can plausibly explain the $\sim$8~Gyr Q-branch cooling delay given a X(\Ne{}) mass fraction of 0.035, a value which is enhanced compared to the expected abundance but considerably lower than the mass fraction of X(\Ne{}) = 0.06 required by \citet{Camisassa21} for sedimentation alone and is consistent with expectations for white dwarf progenitors with high alpha element abundances \citep{Bauer20}.

Assuming a more standard white dwarf composition (X(O) = 0.6, X(\Ne{}) = 0.014), \citet{Blouin21} predict that the cooling delay caused by phase separation will not begin at the onset of core crystallisation, but will instead begin once $\sim$60\% of the core has already crystallised. As a result this will produce a much shorter cooling delay ($\sim$1--2~Gyr) and will produce a narrower pile-up of white dwarfs at cooler temperatures than those produced by C/O crystallisation and phase separation alone. Promisingly, the addition of this cooling delay can neatly explain the \boldnew{\citet{Tremblay19}'s unexpectedly} narrow pile-up in the white dwarf luminosity distribution \citep{Blouin20}\boldnew{. \citet{Blouin21} also point out that} an additional cooling delay that occurs once $\sim$60\% of the core is crystallised is clearly visible in the empirical white dwarf $T_{\text{eff}}-\text{mass}$ diagram of \citet[figures 18, 19]{Kilic20}. These results strongly suggest that \Ne{} phase separation is the energy transport mechanism lacked by traditional models \citep{Blouin21}.

The luminosity function of the local white dwarf population provides a powerful test of white dwarf cooling delays experienced during crystallisation, but its interpretation is importantly sensitive to model assumptions. For example, to simulate the observed population of 0.9--1.1 $M_\odot$ white dwarfs \citet{Tremblay19} assume a constant star formation rate over the past 10 Gyr\boldnew{.} The star formation function of the Milky Way is a topic of intense ongoing research and there is not yet a \textit{communis opinio} on its overall form, though evidence for a variable star formation rate appears to be mounting \citep[see e.g.][and references within]{Tremblay14, Fantin19, Mor19, Alzate21}\boldnew{. The magnitude of variability in the star formation rate proposed in these studies would manifest as second-order variations in the white dwarf luminosity function, which could be comparable to the effects of \Ne{};} indeed, \citet{Fleury22} have recently investigated the formation history of massive white dwarfs assuming the \citet{Mor19} star formation function\boldnew{,} and argue that \boldnew{the resulting variation in stellar formation rates renders the $\sim$8~Gyr cooling delay proposed by \citet{Cheng19} unnecessary} to explain the Q-branch.\footnote{\boldnew{However, \citet{Fleury22} do not discuss the implications of their model for the anomalous kinematics found by \citet{Cheng19}.}}

To circumvent such challenging issues it may thus be enlightening to consider crystallising white dwarfs at an individual level, rather than as a population. The most ideal \boldnew{of such test} cases for core crystallisation models would be white dwarfs whose total ages (i.e. inclusive of the pre-white dwarf lifetime) can be externally constrained. This information is inaccessible for isolated white dwarfs, and must therefore be inferred indirectly by virtue of physical association between a crystallising white dwarf and an object of dateable age. Star clusters provide well-defined coeval stellar populations for which age determination is relatively straightforward, so nearby clusters appear to be an obvious place to look for benchmark crystallising white dwarfs. However, the age distribution of nearby clusters unfortunately makes this less than straightforward; the open cluster population is dominated by young associations ($\lesssim$1 Gyr) whose white dwarfs have not yet cooled enough to reach the crystallisation sequence, whereas the globular cluster population is so old ($\gtrsim$10 Gyr) that their white dwarfs are very low-mass and thus crystallise coincident with convective coupling, making the effects of core crystallisation challenging to detect unambiguously \citep{Bergeron19, Tremblay19}. Only a small number of open clusters with ages within this range (such as the $\sim$8~Gyr old NGC 6791, \citealt{GarciaBerro10}) are amenable for the direct detection of white dwarf pile-ups and cooling delays that can be compared with models, motivating us to look elsewhere for crystallising white dwarfs whose ages can be externally constrained.

\citet{Holberg13} established the term "Sirius-like system" (SLS) to describe binary and multiple star systems that contain at least white dwarf and a non-degenerate star of spectral type earlier than M.\footnote{\boldnew{The main purpose of this distinction of Sirius-like systems from WD + M-dwarf binaries is that in the latter ``...the white dwarf dominates or is at least competitive with the luminosity of the companion at optical wavelengths" \citep{Holberg13}. For earlier-type companions the brightness contrast with the white dwarf becomes progressively larger, making it more difficult to detect the presence of white dwarfs (especially at close separations).}} Since it can be assumed that the components of a SLS are coeval, it is possible to use the non-degenerate members of Sirius-like systems to constrain the total age of the white dwarf component. Previous studies have used this to estimate the masses of white dwarf progenitors \citep{Catalan08}. In principle, if the age of the system and the white dwarf progenitor lifetime can be precisely constrained it would be possible to empirically measure the crystallisation timescale for white dwarfs in Sirius-like systems. If SLSs containing crystallising white dwarfs can be identified they would provide direct constraints on any cooling delays experienced by individual white dwarfs, allowing for granular insight into the mechanisms of energy transport during crystallisation. However, no Sirius-like systems containing white dwarfs undergoing crystallisation have previously been identified. This is undoubtedly due to the difficulty of detecting white dwarfs in SLSs rather than actual absence; indeed, \citet{Holberg13} were aware of only 98 Sirius-like systems and observed a precipitous drop in the number density of known systems beyond 20 parsecs, suggesting that many SLSs remained to be discovered even in nearby space.

In this work we present the discovery of a new Sirius-like quadruple system at 32 parsecs distance, composed of a crystallising white dwarf companion to the previously known triple \thisstar{}. By virtue of its association with these main sequence companions this is the first crystallising white dwarf whose total age can be externally constrained, a fact that we make use of by attempting to empirically measure a cooling delay caused by core crystallisation in the white dwarf.

\vspace{-4mm}
\section{Identification} \label{sec:identification}

The discovery of the system discussed in this work was made as part of a \textit{Gaia}-based search for new nearby Sirius-like systems. This is based on the work of \citet{ElBadry21}, who used observational data from \textit{Gaia}~EDR3 \citep{GaiaEDR3} to assemble a sample of over $10^6$ visual binaries including a large number of previously unknown Sirius-like systems.

In the assembly of their binary sample \citet{ElBadry21} took steps to reduce contamination by chance alignments by adopting relatively stringent cuts in the matching of astrometric parameters. In particular, the authors adopted a cut on the difference between proper motion measurements in stellar pairs requiring this to be consistent with bound orbits assuming a total system mass of 5 $M_\odot$. This is of significant value for ensuring the fidelity of the binary sample, but also causes many unresolved triples and higher-order multiples to be removed from the catalog. This is because the orbital motion of subsystems unresolved by \textit{Gaia} causes the proper motion difference of the resolved wide pair to appear too large to be explained by Keplerian orbital motion \citep{Clarke20}. For example, the nearby Sirius-like system 171~Puppis is not included in the \citet{ElBadry21} catalog because the \textit{Gaia}~EDR3 proper motion of the white dwarf component VB~3 (WD~0743-336) differs significantly from that of the primary\boldnew{. This is due to} the presence of a close companion \boldnew{orbiting the primary} \citep{Tokovinin12}\boldnew{, not spatially resolved by \textit{Gaia}, that} causes the pair to fail the proper motion cut. This effect was also observed for this system by \citet{Hollands18}, who rectified this by utilising the long-term proper motion from PPMXL \citep{Roeser10}, where the effects of subsystem orbital motion are averaged out. As we are interested in nearby Sirius-like systems regardless of their multiplicity, the loss of unresolved high-order multiples from the sample of \citet{ElBadry21} poses a significant drawback.

Fortunately, however, \citet{ElBadry21} have made the code used to construct their binary catalogue freely available.\footnote{\url{https://zenodo.org/record/4435257}} We make use of this to produce a version of the catalogue with a cut on the binary proper motion difference 10 times broader than that used by \citet{ElBadry21}, allowing us to identify a number of Sirius-like systems that were previously excluded from the catalogue as a result of unresolved orbital motion. Although the relaxation of proper motion constraints will necessarily increase the probability of false positives, the use of proper motion measurements from other sources can be used to determine whether the long-term proper motion of the paired stars are consistent with a bound system.

\boldnew{One of the results from} our modification to the search method of \citet{ElBadry21} \boldnew{is the discovery of} a new Sirius-like quadruple system, \thisstar{}, at 32 parsecs distance. We provide a review of current knowledge of this system in Section~\ref{subsec:study_history} and elaborate on the evidence for this association in Section~\ref{subsec:physical_association}.

\subsection{History of study} \label{subsec:study_history}

The primary of the \thisstar{} system is a $\text{V}=7.7$ G-type star located near to the celestial equator. The star does not appear to have been studied before the 1990s and was not known to be near to the Sun prior to the \textit{Hipparcos} mission \citep{Hipparcos}, which measured a stellar parallax of $\varpi=30.13\pm1.37$ mas. The \textit{Hipparcos} astrometric solution includes an acceleration term which suggests the presence of a massive companion, a result supported by \citet{Makarov05} and \citet{Frankowski07} who observed that the proper motion measured by Tycho-2 \citep{Tycho2} disagrees significantly with the \textit{Hipparcos} solution.

The multiplicity of \thisstar{} was confirmed by \citet{Tokovinin16}, who resolved a stellar companion (\thisstar{}~B) at 0.16" ($\sim$5~AU) using SOAR speckle interferometry \citep{Tokovinin08}, and furthermore reported that the system is a spectroscopic triple based on unpublished data. \citet{Fuhrmann19} analysed a spatially unresolved spectrum of \thisstar{} and concluded that three components can be identified, the primary star with $T_{\text{eff}}\approx5650$~K and two companions with $T_{\text{eff}}\approx3900$~K and $\approx4100$~K respectively. Most recently \citet{Tokovinin20} conducted a joint analysis of radial velocity and speckle imaging data for this system, identifying \thisstar{}~Ab and \thisstar{}~B as $\approx$0.45~$M_\odot$ and $\approx$0.61~$M_\odot$ stars with orbital periods of 251 days and 7.45 years respectively. The authors found that the eccentricities of both orbits are relatively low ($e=0.04$ and $0.20$ respectively) and conclude that the stellar orbits are closely aligned, suggesting a degree of dynamical stability in this system.

The existence of a white dwarf close to \thisstarAB{}, identified as Gaia~EDR3 4237555506083389568, was not known prior to \textit{Gaia} DR2. Although the object in question was detected by SDSS (\citealt{SDSS}, identified as SDSS~J200445.49+010929.0 in \citealt{Alam15}), a survey which has led to the discovery of thousands of white dwarfs \citep[e.g.][]{Kleinman04, Kleinman13}, it was not recognised in searches for white dwarfs in the SDSS footprint. However, the star was identified as a white dwarf with high confidence ($P_{\text{WD}}=0.9955$) by \citet{GentileFusillo19} based on \textit{Gaia} DR2 astrometry and photometry. Subsequently \citet{Tremblay20} obtained a spectrum of the star, confirming its white dwarf nature. 

\subsection{Confirmation of physical association} \label{subsec:physical_association}

\begin{table}
	\centering
	\caption{Basic properties of \thisstarAB{} and \thisstarC{} from \textit{Gaia}~EDR3 \citep{GaiaEDR3}.}
	\label{table:basic_properties}
	\begin{tabular}{lcc}
		\hline
		Parameter & \thisstarAB{} & \thisstarC{} \\
		\hline
		RA & 20:04:46.63 & 20:04:45.49 \\
		Declination & +01:09:21.77 & +01:09:29.21 \\
		Parallax (mas) & $29.340 \pm 0.783$ & $30.911 \pm 0.063$ \\
		RA P.M. (\masyr{}) & $-28.15 \pm 0.80$ & $-0.664 \pm 0.065$ \\
		Dec. P.M. (\masyr{}) & $-30.19 \pm 0.52$ & $-28.166 \pm 0.048$ \\ 
		$G$ magnitude & $7.496$ & $16.721$ \\
		$BP$ magnitude & $7.864$ & $16.916$ \\
		$RP$ magnitude & $6.934$ & $16.320$ \\
		\hline
	\end{tabular}
\end{table}

\begin{table*}
	\centering
	\caption{Proper motion measurements of \thisstar{} A. Anomalies are relative to the \textit{Gaia}~EDR3 proper motion of \thisstarC{} ($-0.664 \pm 0.065, -28.166 \pm 0.048$ \masyr{}).}
	\label{table:proper_motions}
	\begin{tabular}{lccccc}
		\hline
		Parameter & \textit{Gaia}~EDR3 & \textit{Hipparcos} & Hipparcos-Gaia & Tycho-2 & PPMXL \\
		\hline
		RA proper motion $\mu_{\text{RA}}$ (\masyr{}) & $-28.15 \pm 0.80$ & $+10.79 \pm 1.96$ & $-1.57 \pm 0.06$ & $-0.6 \pm 0.9$ & $-0.2 \pm 0.9$ \\
		Declination proper motion $\mu_{\text{Dec}}$ (\masyr{}) & $-30.19 \pm 0.52$ & $-38.34 \pm 1.60$ & $-27.33 \pm 0.03$ & $-25.8 \pm 0.9$ & $-24.8 \pm 1.0$ \\
		\hline
		RA proper motion anomaly (\masyr{}) & $-27.47$ & $+11.43$ & $-0.91$ & $+0.08$ & $+0.46$ \\
		Declination proper motion anomaly (\masyr{}) & $-2.02$ & $-10.16$ & $0.84$ & $+2.36$ & $+3.36$ \\
		Tangential velocity anomaly $v_{\text{tan}}$ ($\text{km s}^{-1}$) & $4.22 \pm 0.12$ & $2.36 \pm 0.28$ & $0.19 \pm 0.01$ & $0.39 \pm 0.13$ & $0.54 \pm 0.15$ \\
		\hline
	\end{tabular}
\end{table*}

From our \textit{Gaia}-based search for nearby Sirius-like systems, we identify Gaia~EDR3~4237555506083389568 (hereafter \thisstarC{}) as a possible widely separated quaternary component of the \thisstar{} system. We collate their basic properties from \textit{Gaia}~EDR3 in Table~\ref{table:basic_properties}. The white dwarf lies at a sky separation of 18.3" and at a position angle of $294\degree$ from \thisstarAB{}, equivalent to a projected separation of $\sim$590~AU. The parallaxes of $29.340\pm0.783$~mas and $30.911\pm0.063$~mas for \thisstarAB{} and \thisstarC{} are sufficiently similar to suggest co-distance, although the former is evidently severely perturbed by the close companions \thisstar{}~Ab and B. The \textit{Gaia} parallax of \thisstarC{} establishes a distance of $32.32^{+0.08}_{-0.06}$~pc to the star \citep{BailerJones21}.

The \textit{Gaia}~EDR3 proper motions of \thisstarAB{} and C are sufficiently different that this pair fail the requirements of \citet{ElBadry21} that the tangential velocity difference of a stellar pair must be consistent with bound orbital velocities. Additional scrutiny is therefore required to confirm that the white dwarf is actually a bound component of the \thisstar{} system. The presence of unresolved stellar companions to the primary star provides a plausible explanation for the disagreement in the \textit{Gaia} proper motions, as the orbital reflex velocity induced by the companions would displace the proper motion of \thisstarAa{} from that of the system barycentre over the 2.8-year duration of \textit{Gaia}~EDR3 \citep{GaiaEDR3astrometry}. Evidence that the stellar companions have influenced the astrometry of \thisstarA{} (or, rather, \thisstarAB{}) is provided by the high excess noise of the \textit{Gaia}~EDR3 astrometric solution; the Renormalised Unit Weight Error (RUWE) for this source is 34.676, far above the limit of $>$1.4 for poor astrometric fits suggested by \citet{Lindegren18}. This suggests that the sky motion of \thisstarAB{} is significantly non-linear over the span of \textit{Gaia} observations, such that the proper motion is unlikely to be accurate.

If the stellar companions have displaced the \textit{Gaia} proper motion of \thisstarAB{} away from a barycentric value similar to the proper motion of \thisstarC{}, it would be expected that proper motion measurements from other sources may be in closer agreement with the proper motion of the white dwarf. For this purpose, in Table~\ref{table:proper_motions} we provide a comparison of different proper motion measurements for \thisstarAB{} and their relative differences from the \textit{Gaia}~EDR3 proper motion for \thisstarC{}. We make use of proper motion measurements from \textit{Gaia}~EDR3 \citep{GaiaEDR3}, \textit{Hipparcos} \citep{Hipparcos, HipparcosNew}, the time-averaged Hipparcos-Gaia proper motion \citep{Brandt18, Brandt21}, and the long-timespan measurements from Tycho-2 \citep{Tycho2} and PPMXL \citep{Roeser10}.

Starting with the short-duration astrometry, the \textit{Gaia}~EDR3 proper motions of \thisstarAB{} and C differ by almost 30 \masyr{}. In comparison the anomaly of the \textit{Hipparcos} proper motion is approximately half as large and is furthermore in significant disagreement with the EDR3 measurement; as the observational duration for both \textit{Gaia} and \textit{Hipparcos} are shorter than the 7.45-year orbital period of \thisstar{}~B \citep[$\sim$3.36 yr and $\sim$2.76 yr respectively;][]{Hipparcos, GaiaEDR3astrometry}, it can be inferred that these measurements sample different phases of the stellar orbit.

Next we consider the long-timespan proper motion measurements. The Hipparcos-Gaia proper motion reflects the average motion between the sky positions of \thisstarAB{} measured by \textit{Hipparcos} and \textit{Gaia}~EDR3, and thus provides a proper motion averaged over the $\sim$25 year interval between the missions \citep{Brandt18}. This proper motion agrees remarkably well with that of \thisstarC{}, with a total difference of $\sim$1.2 \masyr{}. This should be taken with a degree of caution since the Hipparcos-Gaia proper motion is dependent on the quality of the \textit{Gaia}~EDR3 astrometry of \thisstarAB{}, which may be significantly inaccurate owing to the very high RUWE of the astrometric solution. However, the Hipparcos-Gaia values for long-term proper motion of \thisstarAB{} is strongly supported by the Tycho-2 and PPMXL data, which differ from the Hipparcos-Gaia proper motion by no more than a few \masyr{} and likewise are in close agreement with the \textit{Gaia}~EDR3 proper motion of \thisstarC{}.

Assuming a mass sum of $2.04~M_{\odot}$ for \thisstarAB{} from \citet{Tokovinin20} and a mass of $0.82~M_{\odot}$ for \thisstarC{} (see Section~\ref{subsec:WD_parameters}), we estimate a total system mass of $2.86~M_{\odot}$. At the 590~AU projected separation between AB and C, the resulting escape velocity for the system is $v_{\text{esc}}=2.9~\text{km s}^{-1}$. As the true separation between the components may be larger than the projected separation this escape velocity should be understood as an upper limit, and likewise the tangential velocity anomalies given in Table~\ref{table:proper_motions} are lower limits for the true orbital velocity. Nevertheless, the tangential velocity anomalies for the long-timespan proper motion measurements are all significantly lower than the system escape velocity ($v_{\text{tan}}\ll1~\text{km s}^{-1}$). Thus, while the association of \thisstarAB{}-C cannot be securely confirmed from the \textit{Gaia} astrometry alone due to the astrometric perturbations from the inner triple system, data from earlier long-timespan proper motion surveys are fully consistent with association of \thisstarC{}. We have therefore discovered that \thisstar{} is a quadruple star and a Sirius-like system.

\vspace{-4mm}
\section{Analysis} \label{sec:analysis}

Having confirmed the physical association of the inner triple \thisstarAB{} with the distant white dwarf \thisstarC{}, we now aim to explore the system in detail. We present an analysis of the physical parameters of \thisstarC{} in Section~\ref{subsec:WD_parameters}, in which we identify the white dwarf as undergoing core crystallisation. Following this discovery, we attempt to constrain the age of the \thisstar{} system in Section~\ref{subsec:system_age} using a variety of techniques. We summarise these results in Section~\ref{subsec:age_summary} and compare the resulting system age with the white dwarf age to test whether there is an observable delay in the cooling of the white dwarf.

\subsection{White dwarf parameters} \label{subsec:WD_parameters}

\subsubsection{Atmosphere model} \label{subsec:atmosphere_model}

\begin{table*}
	\centering
	\caption{Physical parameters of \thisstarC{}.}
	\label{table:WD_parameters}
    \resizebox{\textwidth}{!}{
	\begin{tabular}{lcccccc}
		\hline
		Parameter & \citetalias{GentileFusillo19} & \citet{Tremblay20} & \citet{Tremblay20} & \citet{McCleery20} & \boldnew{\citetalias{GentileFusillo21}} & This work \\
		  & Photometry & Spectroscopy & Photometry & Photometry & \boldnew{Photometry} & Photometry \\
        & \boldnew{(\textit{Gaia} DR2)} &  & \boldnew{(\textit{Gaia} DR2)} & \boldnew{(\textit{G.} DR2 + PS $grizy$)} & \boldnew{(\textit{Gaia} EDR3)} & \boldnew{(\textit{G.} EDR3 + PS $giz$)} \\
		\hline
		$T_{\text{eff}}$ (K) & $6531 \pm 70$ & $6770 \pm 60$ & $6510 \pm 70$ & $6126 \pm 60$ & \boldnew{$6642\pm78$} & $6600 \pm 80$ \\
		$\log g$ (${\rm cm}^2{\rm s}^{-1}$) & $8.325 \pm 0.029$ & $8.30 \pm 0.10$ & $8.32 \pm 0.03$ & $8.076 \pm 0.028$ & \boldnew{$8.366\pm0.030$} & $8.353 \pm 0.028$ \\
		Mass ($M_\odot$) & $0.798 \pm 0.019$ & -- & -- & -- & \boldnew{$0.825\pm0.020$} & $0.817 \pm 0.019$ \\
		\hline
	\end{tabular}
    }
\end{table*}

\begin{figure*}
	\includegraphics[width=\columnwidth]{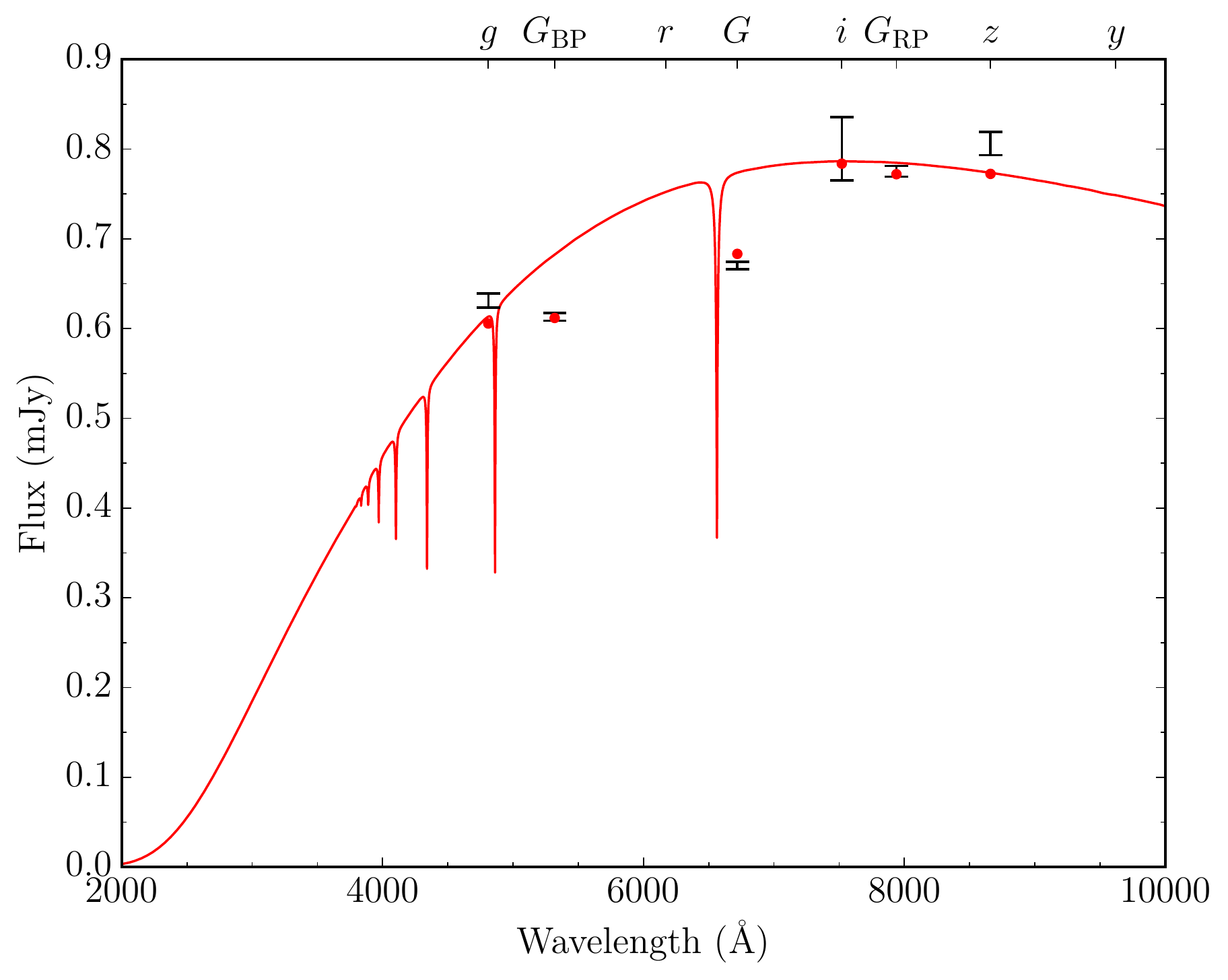}
	\includegraphics[width=\columnwidth]{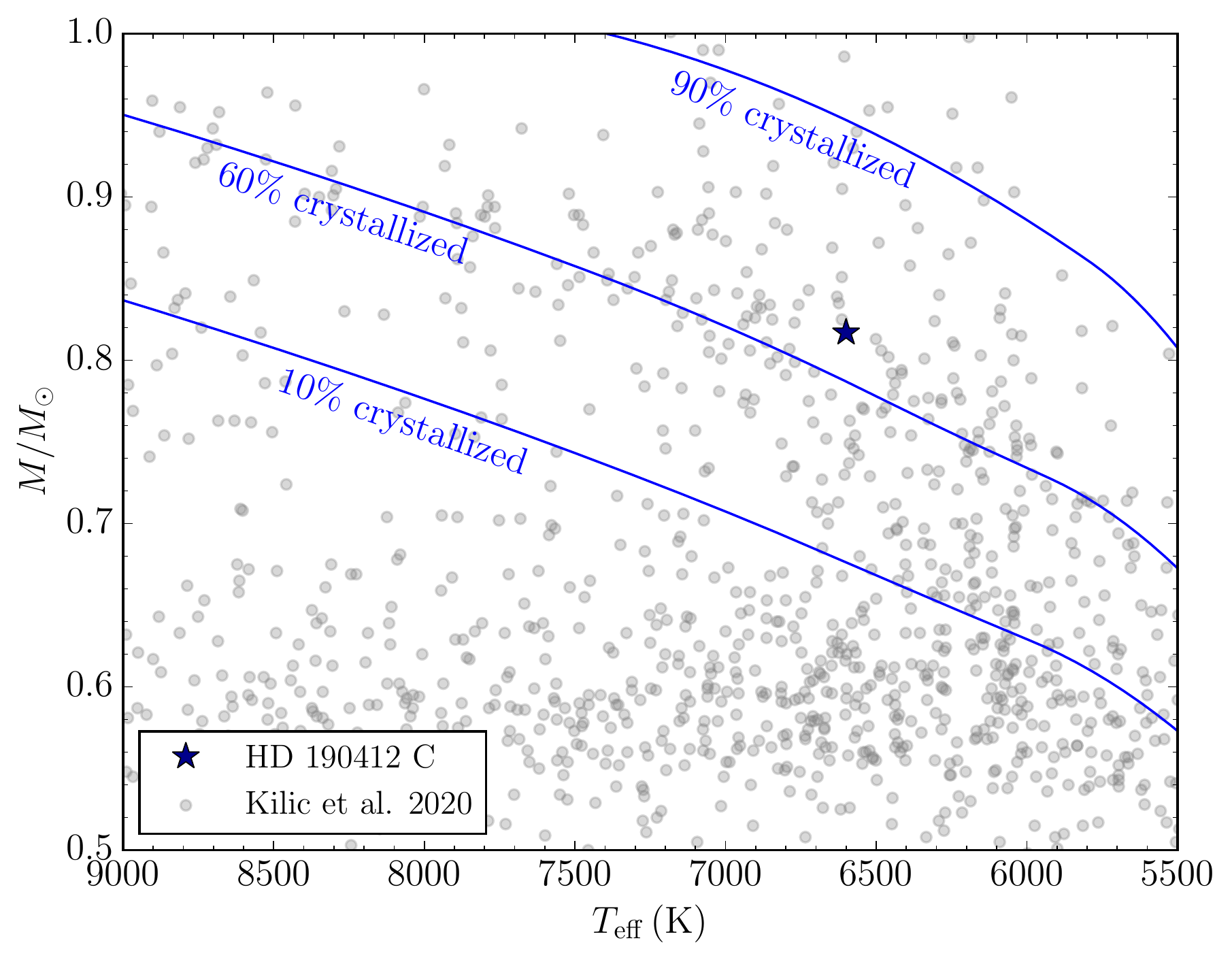}
    \caption{(Left) Photometric fit of the \textit{Gaia}~EDR3 and Pan-STARRS photometry of \thisstarC{}. The error bars represent the observational data and the synthetic fluxes are in red. The Pan-STARRS $r$ and $y$ bands both yield implausible fluxes of over 1\,mJy and were ignored for our analysis. (Right) Mass and effective temperature of \thisstarC{} (star) and of white dwarfs in the \citet{Kilic20} sample (circles). The blue lines indicate where 10\%, 60\%, and 90\% of the core is crystallised assuming an homogeneous core composition with $X({\rm O})=0.60$. There is a clear pile-up of objects corresponding to $\approx60\%$ core crystallisation, which \citet{Blouin21} suggests to be the result of a cooling delay caused by \Ne{} phase separation. As \thisstarC{} lies within this overdensity in the T$_{\text{eff}}-\text{mass}$ plane, it is an important benchmark for understanding this feature of the white dwarf population.}
    \label{figure:WD_fit}
\end{figure*}

To evaluate the physical properties of the white dwarf \thisstarC{}, we first need to determine its atmospheric parameters. To fully take advantage of the high-precision \textit{Gaia} astrometry we make use of the photometric technique \citep[see][]{Bergeron97,Bergeron19}. We use the atmosphere models described in \citet[and references therein]{Blouin18a,Blouin18b}, the parallax measurement from \textit{Gaia}~EDR3, and photometry from \textit{Gaia}~EDR3 and Pan-STARRS. We found that only the $g$, $i$, and $z$-band Pan-STARRS photometry could be used because the $r$ and $y$ fluxes diverge strongly from the expected spectral energy distribution, most likely because of flux contamination by \thisstarAB{}. The photometry used in our model is provided in Appendix~\ref{appendix:WD_photometry}. \citet{Tremblay20} observed the spectrum of \thisstarC{} and found that it has a hydrogen-dominated (DA-type) atmosphere; we thus assume a pure-hydrogen atmospheric composition for our fit to the photometry.

Our best-fit model \boldnew{for the spectral energy distribution (SED)} is shown in the left panel of Figure~\ref{figure:WD_fit}, and the corresponding physical parameters are given in Table~\ref{table:WD_parameters} along with values from previous studies for the purpose of comparison. Our solution is in good agreement with the parameters obtained by \citet{GentileFusillo19} and \citet{Tremblay20} based on fits to the \textit{Gaia}~DR2 photometry, \boldnew{as well as} the spectroscopic parameters from the latter study based on a fit to the Balmer lines. \boldnew{We also find good agreement with the more recent parameters from \citet{GentileFusillo21} based on the \textit{Gaia}~EDR3 photometry.} We note however that the Pan-STARRS-based photometric solution of \cite{McCleery20} differs significantly from all other results; this may be because they included the contaminated $r$ and/or $y$ band photometry in their analysis.

In the right panel of Figure~\ref{figure:WD_fit} we plot the position of \thisstarC{} in the $T_{\text{eff}}-\text{mass}$ diagram along with white dwarfs from the \citet{Kilic20} sample. We observe that \thisstarC{} can be comfortably placed in the theoretically predicted temperature range of C/O core white dwarfs undergoing core crystallisation. This makes this the first crystallising white dwarf belonging to a Sirius-like system to \boldagain{be confirmed}.\footnote{\boldagain{By this we recognise that large binary samples such as \citet{ElBadry21} undoubtedly contain other Sirius-like systems with crystallising white dwarfs, but \thisstarC{} is the first to be individually validated.}} Furthermore, it can also be seen that \thisstarC{} lies in an overdensity of white dwarfs found along the line of $\approx 60\%$ core crystallisation. This pile-up of white dwarfs \boldnew{is visible in the $T_{\text{eff}}-\text{mass}$ diagram of} \citet{Kilic20}, who found that their cooling models could not reproduce this feature. \citet{Blouin21} advanced the hypothesis that a significant cooling delay caused by \Ne{} phase separation could explain this overdensity. The position of \thisstarC{} within this pile-up raises the possibility that it could be a powerful benchmark for testing models of white dwarf cooling and crystallisation. This motivates us to develop a detailed model of the cooling of this white dwarf.

\subsubsection{Cooling model} \label{subsec:cooling_model}

To transform \thisstarC{}'s atmospheric parameters into a cooling age, we use the state-of-the-art techniques provided by the latest version of the STELUM stellar modelling package \citep{Bedard20,Bedard22}. Our cooling sequences include the gravitational energy release from C/O phase separation \citep[using the phase diagram of][]{Blouin21PRE} and the gravitational settling of \Ne{} in the liquid phase (using diffusion coefficients discussed below). However, the distillation mechanism associated with \Ne{} phase separation is currently not included in STELUM. We assume a canonical $10^{-2}\,M_{\star}$ He envelope and a ``thick'' H envelope of $10^{-4}\,M_{\star}$, which is justified by the DA nature of \thisstarC{} \citep[e.g.,][]{Renedo10}. For simplicity, we assume an homogeneous C/O core with an O mass fraction of $X({\rm O})=0.60$. This is a first-order approximation of the predictions of stellar evolution codes \citep{Salaris10,Althaus12,Bauer20}. We refrain from using a more complex C/O composition profile given the high level of uncertainty surrounding the core composition profile of white dwarfs \citep{Salaris17,Giammichele18,Giammichele22,DeGeronimo19}. We also include a uniform trace of \Ne{} in the core with $X(\Ne{})=0.008$. This value was chosen because it corresponds to the metallicity of the system ($\text{[M/H]}=-0.24$, \citealt{Brewer16}), and the \Ne{} content of a white dwarf corresponds in very good approximation to the metallicity of its progenitor \citep{Cheng19,Salaris22}.

\begin{figure}
	\includegraphics[width=\columnwidth,clip=true,trim=2.5cm 7.1cm 2.5cm 7.8cm]{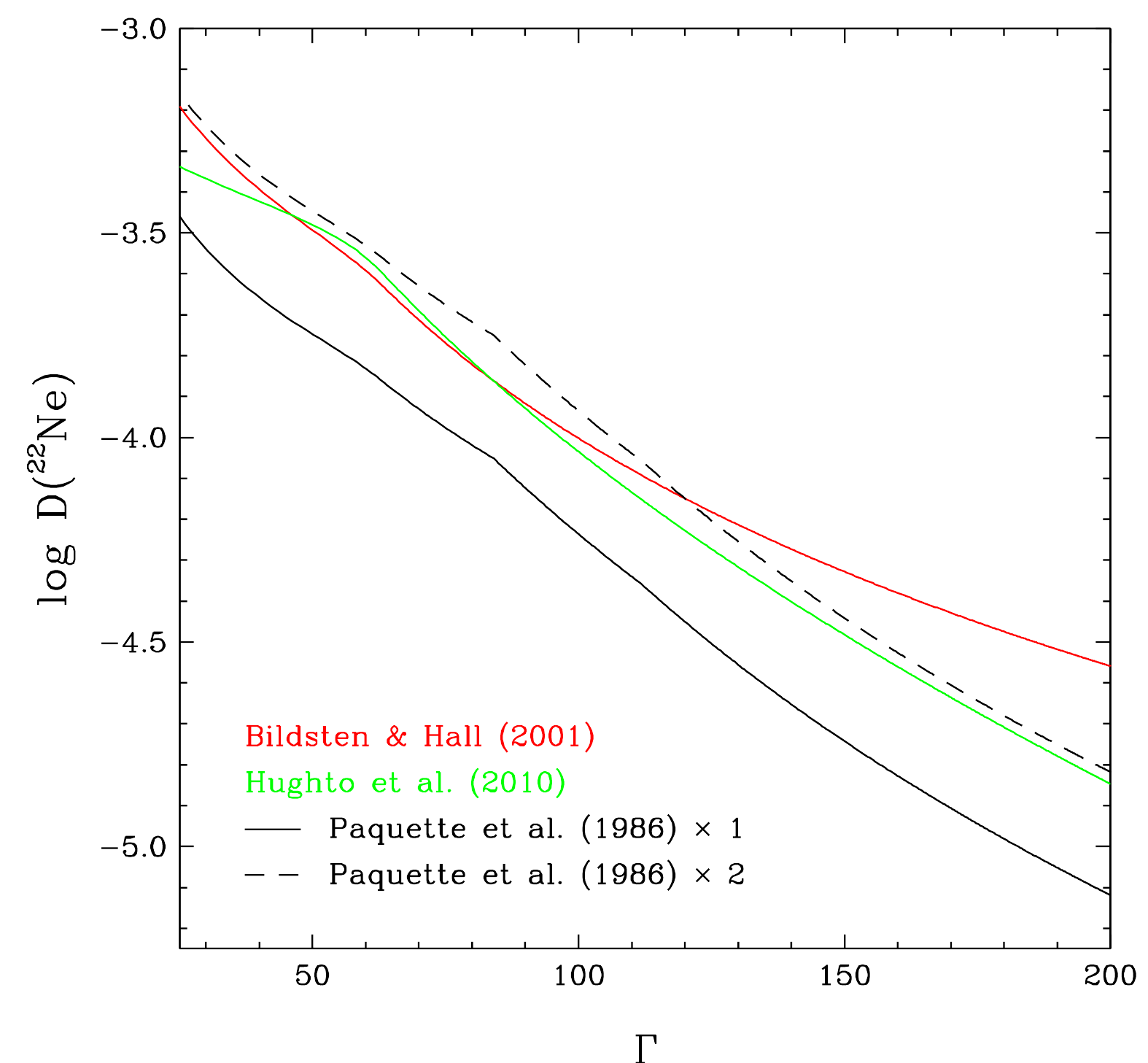}
    \vspace{-3mm}
    \caption{Comparison of various \Ne{} diffusion coefficients as a function of the Coulomb coupling parameter in a white dwarf model with $M = 0.82\,M_{\odot}$ and $T_{\rm eff} = 9500$ K (just before the onset of core crystallisation). The black, red, and green lines correspond to results obtained from the methods of \citet{Paquette86}, \citet{Bildsten01}, and \citet{Hughto10}, respectively. The dashed black line is offset from the solid black line by a factor of 2.}
    \label{figure:diffcoeff}
\end{figure}

The efficiency of \Ne{} gravitational settling and the magnitude of the associated energy release depend on the adopted \Ne{} diffusion coefficient \citep{GarciaBerro08}. STELUM relies on diffusion coefficients computed using the method developed by \citet{Paquette86} and updated by \citet{Fontaine15}, which are valid for moderately coupled plasmas. These can be realistically extended to strongly coupled plasmas (which is the relevant regime here) through a physically motivated extrapolation, which may however introduce mild systematics (see \citealt{Paquette86} for details). Figure~\ref{figure:diffcoeff} shows the diffusion coefficient for \Ne{} as a function of the Coulomb coupling parameter in a $0.82\,M_{\odot}$ model just before the onset of core crystallisation. Also displayed are the coefficients obtained from the approximate analytic expression of \citet{Bildsten01} and from the molecular dynamics simulations of \citet{Hughto10}, the latter arguably being more accurate in the strongly coupled regime (see also \citealt{Caplan22} for similar calculations). Coincidentally, the \citet{Paquette86} coefficient is lower than the \citet{Hughto10} coefficient by almost exactly a factor of 2 over most of the liquid core. Therefore, in our evolutionary calculations, we set the \Ne{} diffusion coefficient to twice that predicted by the method of \citet{Paquette86}. Although it would also be possible to employ the results of \citet{Hughto10} directly, our approach is essentially equivalent and computationally simpler: STELUM models atomic diffusion using the formalism of \citet{Burgers69}, which involves the so-called resistance coefficients $K_{ij}$ rather than the diffusion coefficients $D_{ij}$ (see \citealt{Bedard22} for details). The \citet{Paquette86} method allows a straightforward calculation of the resistance coefficients for all elements, whereas \citet{Hughto10} only provide the \Ne{} diffusion coefficient. Figure~\ref{figure:diffcoeff} indicates that, if anything, our strategy may very slightly overestimate the efficiency of \Ne{} gravitational settling.

\begin{figure}
    \label{figure:evol}
	\includegraphics[width=\columnwidth]{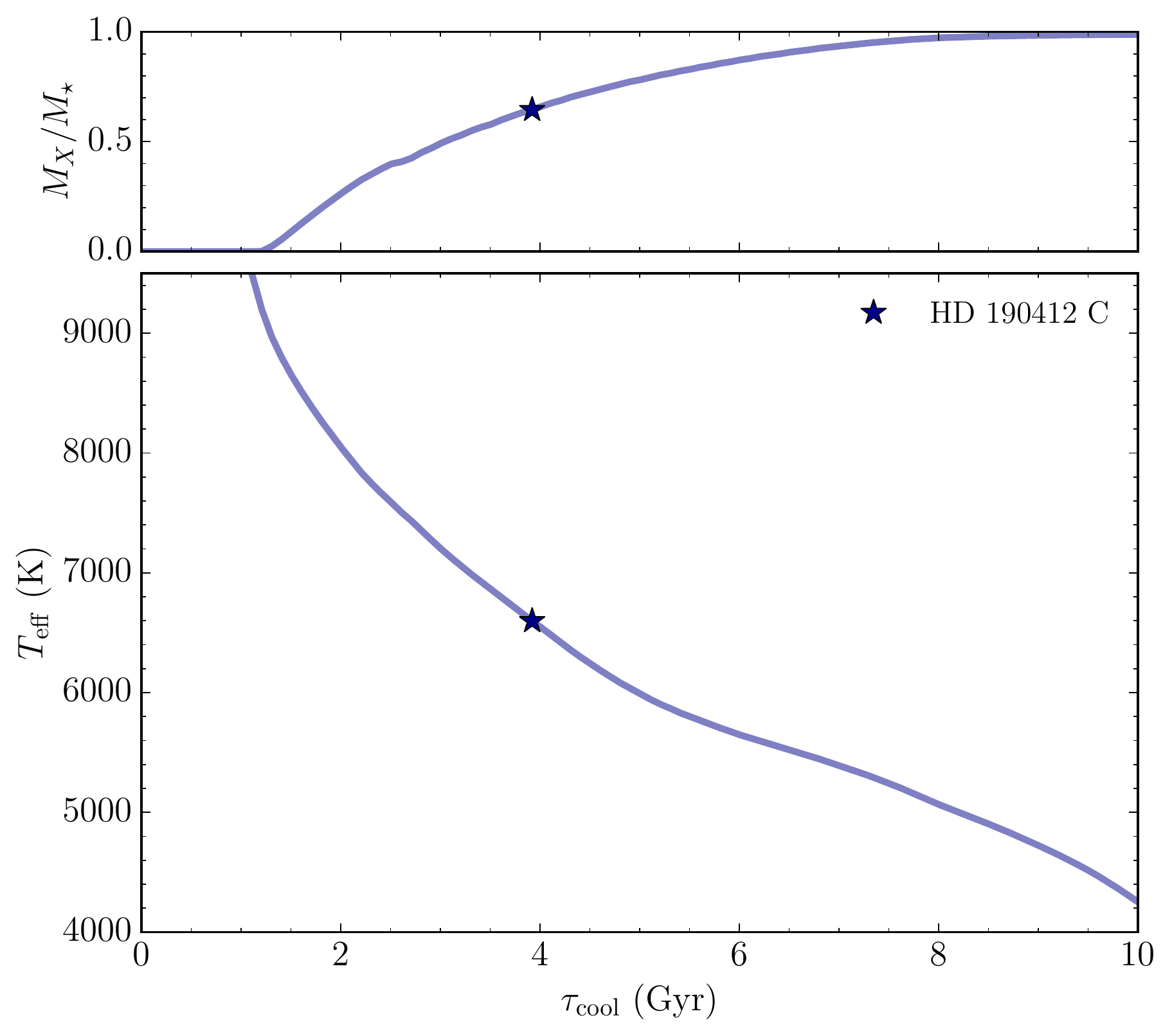}
    \caption{Cooling model of a $0.82\,M_{\odot}$ DA white dwarf (see text for details). The top panel shows the fraction of the C/O core that is crystallised as a function of the cooling age and the bottom panel shows the evolution of the effective temperature. A star indicates the current temperature of \thisstarC{}. We estimate a cooling age of $3.9 \pm 0.2\,{\rm Gyr}$ and a core crystallisation fraction of 65\% for the white dwarf.}
\end{figure}

Figure~\ref{figure:evol} shows the cooling of \thisstarC{} as predicted by our evolutionary calculations. We find a cooling age of $3.9 \pm 0.2\,{\rm Gyr}$, where the confidence interval was obtained by propagating the uncertainties on the mass and effective temperature. The top panel of Figure~\ref{figure:evol} shows that, given our assumption on \thisstarC{}'s core composition, 65\% of its core is expected to be crystallised. \cite{Blouin21} predict that a white dwarf with a homogeneous $X({\rm O})=0.60$ core that evolved from a solar-metallicity progenitor ($X(\Ne{})=0.014$) will undergo phase separation of \Ne{} once $\approx$60\% of the core has crystallised (see Section~\ref{subsec:atmosphere_model}), a process that can cause a substantial delay of the cooling of the white dwarf. We therefore anticipate that \Ne{} phase separation could significantly affect the cooling age of \thisstarC{}. The effect of \Ne{} gravitational settling (which is included in our calculations) on the \Ne{} content of the core is very small; compared to an analogous evolutionary sequence in which diffusion is turned off, the measured cooling age is only 0.15 Gyr larger, and the mass of \Ne{} above the crystallisation front is reduced by only 15\%. This means that there is a significant reservoir of liquid \Ne{} available for the distillation process. We recall that the \Ne{} distillation mechanism is currently not included in STELUM, hence the $3.9 \pm 0.2\,{\rm Gyr}$ cooling age given above does not account for the effects of \Ne{} phase separation and may therefore be underestimated. We return to the effects of \Ne{} phase separation in Section~\ref{subsec:WD_discussion}.

\subsubsection{Progenitor mass and lifetime}  \label{subsec:WD_lifetime}

Having now constrained the physical parameters of the white dwarf \thisstarC{}, we next consider the nature of the progenitor star. A white dwarf generally carries little information about its progenitor, however there is a relationship between the mass of the white dwarf and that of the progenitor star, known as the initial--final mass relation (IFMR). By estimating the progenitor mass it is possible to further estimate the pre-white dwarf lifetime of the progenitor. This is an important step for constraining the total age, as the cooling age of a white dwarf does not account for the pre-white dwarf lifetime of the progenitor star.

To estimate the progenitor mass of \thisstarC{} we use the semi-empirical IFMR of \citet{Cummings18}. Fortunately the white dwarf mass of $0.817\pm0.019~M_\odot$ lies in a densely sampled area of the IFMR, allowing us to precisely constrain the mass of the progenitor. Using the \citet{Cummings18} IFMR, we find a progenitor mass of $3.38^{+0.13}_{-0.10}~M_\odot$; this corresponds to the main sequence mass of a late-B type star. Next, using the MIST theoretical isochrones \citep{Dotter16}, we derive a pre-white dwarf lifetime of $290\pm30$~Myr for \thisstarC{}, assuming a metallicity of $\text{[Fe/H]}=-0.25$ from \citet{Brewer16}.

As a sanity check, we compare our results \boldnew{to} WD~0833+194, a white dwarf member of the Praesepe cluster with a similar mass to \thisstarC{} ($0.813\pm0.027~M_\odot$). \citet{Cummings18} estimate a white dwarf cooling age of $364^{+33}_{-30}$~Myr and a cluster age of $685\pm25$~Myr based on MIST isochrones. This implies a progenitor lifetime of $\approx$321~Myr for WD~0833+194, which agrees well with our value for \thisstarC{}. The authors further report a progenitor mass of $3.51^{+0.12}_{-0.10}~M_\odot$ for this white dwarf, within 1.3$\sigma$ of our estimate. Small differences between our results can perhaps be ascribed to differences in metallicity, as Praesepe is significantly more metal-rich than \thisstar{} ($\text{[Fe/H]}=0.15$, \citealt{Cummings18}). We thus conclude that our values of the progenitor mass and lifetime for \thisstarC{} are physically reasonable.

Combining the $3.9\pm0.2$~Gyr cooling age with the $0.29\pm0.03$~Gyr pre-white dwarf lifetime, we estimate a total age of $4.2\pm0.2$~Gyr for \thisstarC{}. As previously noted, this value does not account for the effects of \Ne{} phase separation, which we expect would result in a significantly higher cooling age for this star. The association of \thisstarC{} with the main sequence stars \thisstarAB{} makes this the first \boldagain{confirmed} crystallising white dwarf for which the total age can be constrained, by assuming coevality for all members of the system. This means that it is possible in principle to empirically detect any additional delays in the cooling of \thisstarC{} not accounted for in our cooling model, thus potentially allowing for a direct constraint on the cooling delay caused by \Ne{} distillation. For this purpose we therefore turn to estimation of the system age.

\subsection{System age} \label{subsec:system_age}

Having identified \thisstarC{} as a white dwarf undergoing core crystallisation and measured its fundamental parameters and cooling age, we now aim to measure its total lifetime through its companions \thisstarAB{}. Unfortunately, measuring the age of main-sequence stars is a challenging endeavour, far more so than for white dwarfs. In the words of \citet{Mamajek08a}, to ask what is the age of a particular star ``is probably one of the most frustrating astronomical inquiries that one can make, and also one of the most quixotic of tasks to tackle.'' Nevertheless, techniques for measuring the ages of field stars have seen a great deal of interest and development in recent years, and we aim to make use of as many different methods as possible to constrain the age of the \thisstar{} system. In the following sections we describe each of these methods and their resulting age constraints in turn.

\subsubsection{Isochronal age}

\begin{figure}
	\includegraphics[width=\columnwidth]{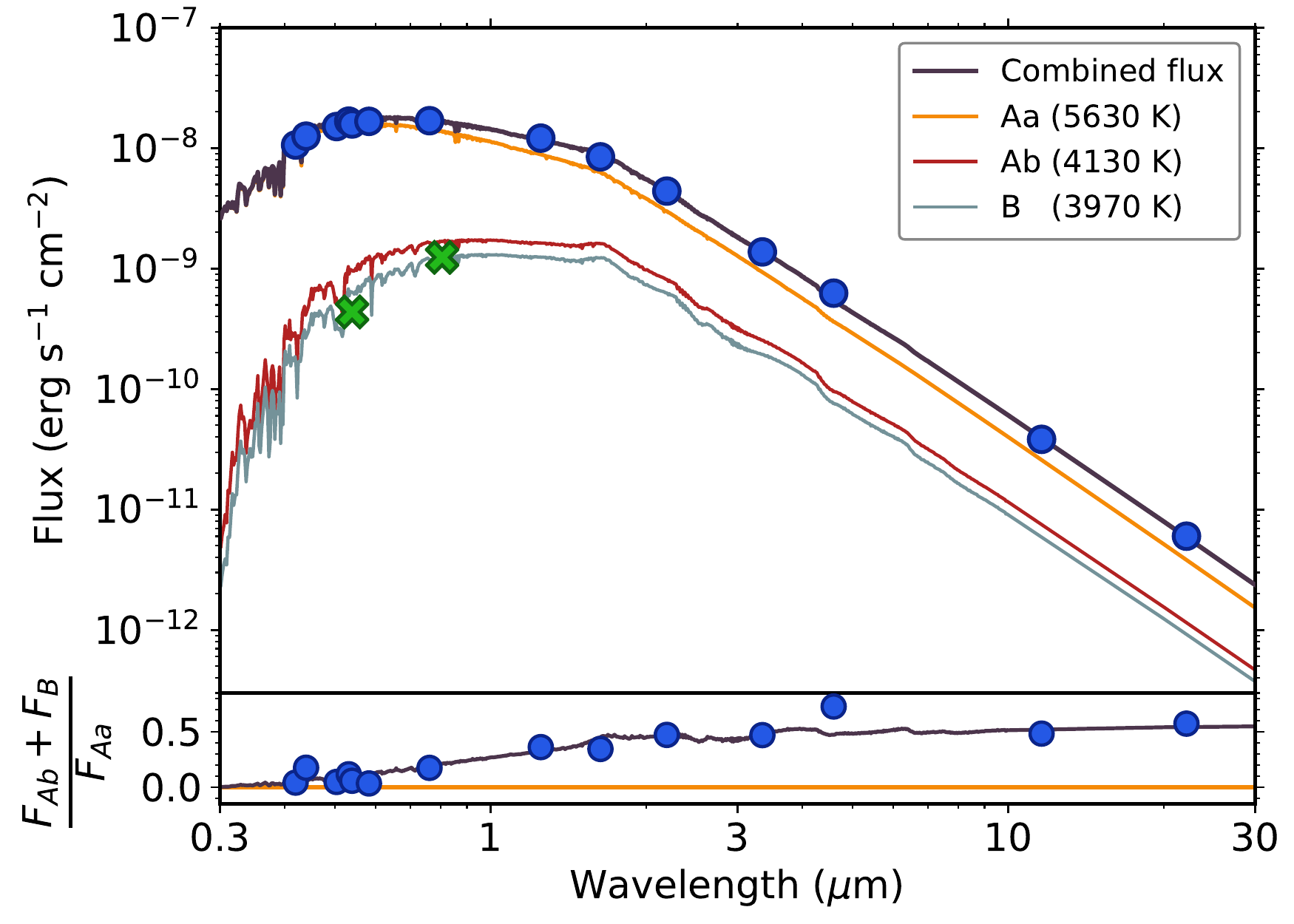}
    \vspace{-6mm}
	\caption{\boldnew{The SED of \thisstarAB{}. We use the ATLAS9 stellar models \citep{ATLAS9} to show the flux contributions of three stars in the inner triple. The points in blue are the spatially unresolved photometry, while the two points in green are the resolved flux of \thisstar{}~B (see Appendix~\ref{appendix:isochrones}). While the flux from Ab and B is negligible in the visible, in the infrared their emission grows as high as 50\% of that from A.}}
	\label{figure:SED_MS}
\end{figure}

\begin{figure*}
	\includegraphics[width=\textwidth, clip, trim=0 0.8cm 0 0]{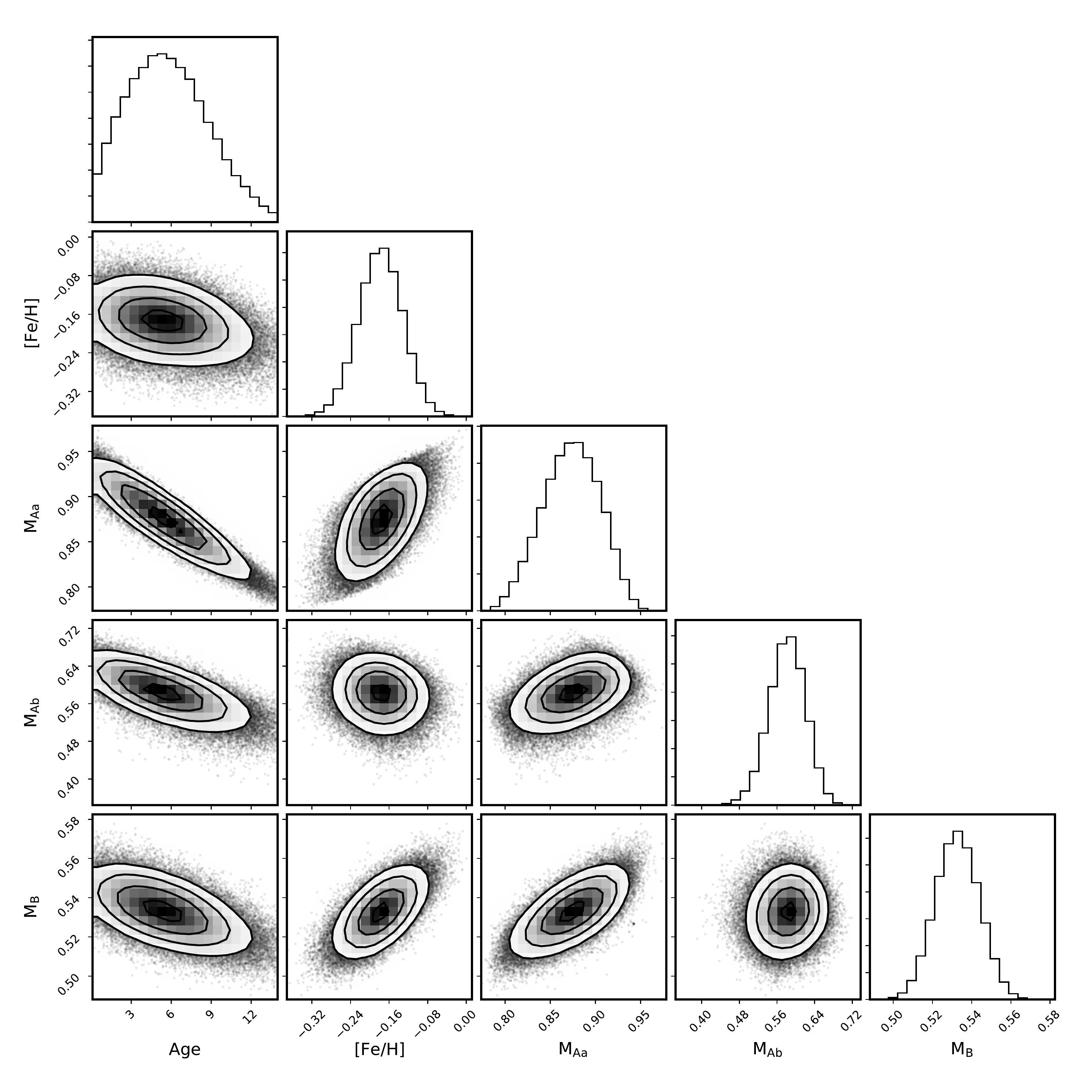}
	\caption{Joint posterior distributions and histograms of the posteriors for our isochrone fit. We find a best-fit system age of $5.6^{+3.3}_{-3.0}$~Gyr and masses of $M_{\text{Aa}},\,M_{\text{Ab}},\,M_{\text{B}}=0.87\pm0.03,\,0.58\pm0.04,\,0.533\pm0.011~M_\odot$ for the three stars in the model, as well as an [Fe/H] of $-0.18\pm0.05$.}
	\label{figure:age_isochrone}
\end{figure*}

Isochrone fitting is a commonly used method for estimating stellar ages, being perhaps the method used most frequently for field stars. The basic principle of the technique is that the observed temperature and luminosity of a star can be reproduced by a model isochrone that assumes a set of physical parameters (e.g. mass, age, metallicity). Typically the variable parameters other than mass and age can be constrained by external methods, so in the presence of sufficiently precise observational data it is possible to extract the age of a star from an isochrone fit.

However, there are various complications that make isochronal age measurement challenging. Among the most important is that isochrones tend to be largely invariant across the main sequence since stellar evolution is slow at this evolutionary stage, such that main sequence stars may be consistent with model isochrones spanning billions of years in age within measurement uncertainties \citep{Angus19}. This is indeed true for \thisstarAa{}, which is a G-type dwarf star. Another complication that is relevant for \thisstarAB{} in particular is that this is a triple star unresolved in most forms of observation. Though the lower-mass components Ab and B are much less luminous than \thisstarAa{}, their lower temperature means that their contribution to the stellar flux grows more significant towards longer wavelengths, and this flux must be accounted to produce accurate results in isochrone fitting \citep{VallsGabaud14}. This amounts to modelling the magnitudes of each component individually and then calculating their combined magnitude from the sum of their fluxes, which can be calculated as

\begin{equation}
m_\text{tot}=-2.5\log_{10}\left(10^{-0.4m_{\text{Aa}}}+10^{-0.4m_{\text{Ab}}}+10^{-0.4m_{\text{B}}}\right)
\:,
\end{equation}

where $m_{\text{Aa}}$ is the photometric magnitude of \thisstarAa{}, and so forth. The model magnitude sum $m_\text{tot}$ can then be fitted against the observed magnitudes.

In this study we use the MIST isochrones \citep{Dotter16, Choi16} to model the photometry of \thisstarAB{}. For input data we use spatially unresolved photometry of \thisstarAB{} from the literature, while we employ a series of priors on the system and stellar parameters to ensure that our results are physically realistic. This information is detailed in Appendix~\ref{appendix:isochrones}. We use \texttt{emcee} \citep{emcee} to sample the parameter space of our model. As variable parameters we use distance, age, [Fe/H], and masses for each component of the inner triple. As \thisstar{} is relatively nearby (32 parsecs), we assume that there is no reddening or extinction of the photometry for our fit.

\boldnew{We visualise the SED resulting from our isochrone fit to \thisstarAB{} in Figure~\ref{figure:SED_MS}. We find a} best-fit system age of $5.6^{+3.3}_{-3.0}$~Gyr\boldnew{, and effective temperatures for the three components of $T_{\text{eff,Aa}},\,T_{\text{eff,Ab}},\,T_{\text{eff,B}}=5630\pm30,\,4130\pm150,\,3970\pm50$~K.} The component masses are $M_{\text{Aa}},\,M_{\text{Ab}},\,M_{\text{B}}=0.87\pm0.03,\,0.58\pm0.04,\,0.533\pm0.011~M_\odot$, resulting in a total mass of $M_{\text{tot}}=1.99\pm0.07~M_\odot$ that agrees well with the dynamical prior of $2.06^{+0.13}_{-0.12}~M_\odot$ (Appendix~\ref{appendix:isochrones}). \boldnew{Their respective luminosities are $0.68\pm0.02,\,0.083^{+0.025}_{-0.020},$ and $0.057\pm0.003~L_\odot$. While Ab and B contribute a negligible amount of flux in visible wavelengths, towards the infrared their combined emission reaches as high as 50\% of the flux of the primary. Our model iron abundance of $[\text{Fe/H}]=-0.18\pm0.05$ is slightly (1$\sigma$) higher than our prior value of $-0.23\pm0.05$.}

We visualise the posteriors of our model using \texttt{corner} \citep{corner} in Figure~\ref{figure:age_isochrone}. The model age is most strongly correlated with the mass of \thisstarAa{}, with higher masses leading to lower ages. The age posterior is relatively Gaussian, \boldnew{but is unsurprisingly very loosely constrained --} the width of the distribution spans the entire prior \boldnew{range} of $0.1-14$~Gyr, clearly indicating that it is challenging to precisely constrain the age of this system with isochrones.

\subsubsection{Kinematic age}

It has long been known that the motion of a star through the galaxy is related to its age to a significant degree, with older stars generally possessing faster space velocities and vice versa. \citet{Gilmore83} were the first to propose that the Milky Way disk could be split into "thin" and "thick" components, with the thick disk characterised by a larger scale height and age than the thin disk as well as differences in chemical abundance, a theory which is now widely accepted. The thick disk is generally agreed to be somewhat older than the thin disk and had ceased star formation by approximately $\sim$8~Gyr ago, whereas the thin disk is thought to have begun to form $\approx$8-10~Gyr ago and has continued forming stars until the present \citep{Fuhrmann11, Xiang17}. The best technique to identify the disk membership of a star is through its kinematics; several studies have also employed $\alpha$-element abundances for this purpose, however \citet{Hayden17} argue that this can be misleading. We approach $\alpha$-element abundances in a different way in Section~\ref{subsec:chemical_clocks}.

We calculate the space velocities of \thisstarAB{} following the method of \citet{Johnson87}. We assume the stellar co-ordinates from \textit{Gaia}~EDR3 and a system radial velocity of $-55.34\pm0.03$~km~s$^{-1}$ from \citet{Tokovinin20}. For the system proper motion we use the Tycho-2 values, which we assume are not strongly affected by the orbital motion of \thisstar{}~Ab and B (see Section~\ref{subsec:physical_association}). Finally, for the parallax we use the \textit{Gaia}~EDR3 measurement for \thisstarC{}, as the parallax for A is affected by orbital motion. With these constraints, we derive space velocities of $(U,V,W)=(-37.21\pm0.10, -38.97\pm0.11, 13.23\pm0.13)$~km~s$^{-1}$ for \thisstarAB{}. Compared to the stellar sample of \citet{Ramirez07}, \thisstarAB{} lies within the velocity range of high-probability thin disk members. We therefore conclude that the \thisstar{} system belongs to the thin disk. This provides us with a kinematic upper limit on the system age of <10~Gyr, since as previously discussed the oldest members of the thin disk are thought to be no older than this.

Recent studies have demonstrated that further constraints on stellar ages can be derived from kinematics. \citet[][hereafter \citetalias{AlmeidaFernandes18}]{AlmeidaFernandes18} presented a novel method for estimating stellar ages based on kinematic evidence calibrated for thin disk stars by modelling the velocity dispersion of stars against their isochronal ages based on data from the Geneva-Copenhagen Survey \citep{Nordstrom04, Casagrande11}. Owing to the statistical nature of the method the resulting age estimates are necessarily imprecise, with a median 1$\sigma$ uncertainty of $\approx$3~Gyr. However, the authors reason that kinematic age estimates are useful for stars that cannot be precisely aged using isochrones, which is indeed the case for \thisstar{}.

\citet[][hereafter \citetalias{Veyette18}]{Veyette18}, writing shortly after the publication of \citetalias{AlmeidaFernandes18}, offered some revisions on their method. In particular, the authors note that the Geneva-Copenhagen Survey sample is biased towards brighter, and hence more massive, solar-type stars. As a result young stars are over-represented in the sample used by \citetalias{AlmeidaFernandes18}, which could introduce biases in their kinematic age estimates. To counteract this \citetalias{Veyette18} introduced a mass cut to their sample selection ($0.9<M_*<1.1~M_\odot$), which resulted in a stellar age distribution significantly less skewed towards young stars.

\begin{figure}
	\includegraphics[width=\columnwidth]{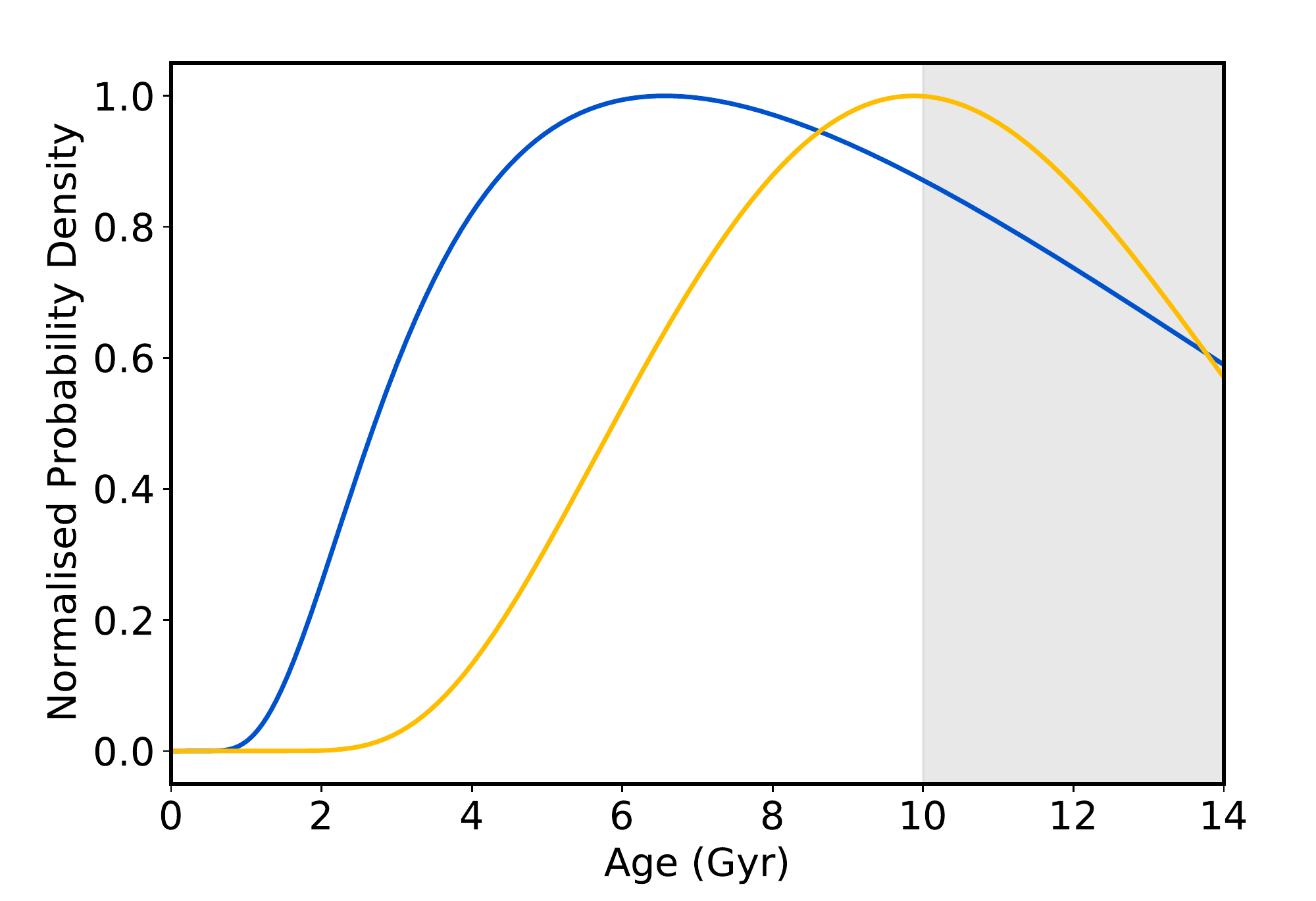}
	\caption{The normalised probability density function for the age of the \thisstar{} system based on the \textit{UVW}-age relationships of \citetalias{AlmeidaFernandes18} (blue) and \citetalias{Veyette18} (gold). We derive 1$\sigma$ confidence intervals on the system age of $7.8^{+3.8}_{-3.5}$~Gyr and $9.6^{+2.7}_{-2.9}$~Gyr respectively. Ages above >10~Gyr, which we consider to be implausible for a member of the thin disk, are shaded in grey.}
	\label{figure:age_kinematic}
\end{figure}

To derive the kinematic age of \thisstar{}, we apply the $UVW$-age relationships of \citetalias{AlmeidaFernandes18} and \citetalias{Veyette18} to the space velocities calculated above. We plot the resulting normalised probability density functions from both relations in Figure~\ref{figure:age_kinematic}. The age distributions are spread unsurprisingly broadly in both cases, but agree reasonably well with each other and suggest a relatively old age for the system. We find most probable ages of 6.6~Gyr and 9.9~Gyr and 1$\sigma$ confidence intervals of $7.8^{+3.8}_{-3.5}$~Gyr and $9.6^{+2.7}_{-2.9}$~Gyr from the two $UVW$-age relationships respectively. We note that \citetalias{AlmeidaFernandes18} explicitly exclude thick disk stars from their sample but set an upper limit for their age distribution of only <14~Gyr, allowing for >10~Gyr ages that we would consider to be implausibly large for thin disk stars. We therefore propose an upper limit of <10~Gyr as an ancillary kinematic constraint on the system age alongside the results from the $UVW$-age relationships.

\subsubsection{Chemical clock age} \label{subsec:chemical_clocks}

In recent years, there have been a great number of studies focused on the relationship between stellar elemental abundances and age. This interest began in earnest with \citet{Nissen15}, who studied the abundances of a sample of 21 solar twins and found that ``for several elements there is an astonishingly tight correlation between [X/Fe] and stellar age.'' In particular, the author observed that the relative abundances of the $\alpha$-process elements generally increase with the isochronal age of the star, with [Mg/Fe] showing a particularly strong correlation, while the $s$-process element yttrium shows an anti-correlation between age and [Y/Fe]. Making use of these opposing correlations, \citet{Nissen15} found that stellar [Y/Mg] abundances are especially strongly correlated with age, a result which was further supported by \citet{TucciMaia16} who suggested that the age-[Y/Mg] could be used to determine stellar ages to $\sim$0.8~Gyr precision.

The universality of such "chemical clocks" was questioned by \citet{Feltzing17}, who observed that the preceding studies were restricted to stars of Sun-like metallicities. Extending their sample to stars with a broader range of [Fe/H] abundances, the authors found that the age-[Y/Mg] relation varies for different metallicities. This result was not recovered by \citet{Titarenko19}, however \citet{Casali20} found similar metallicity-based variance. This issue was considered extensively by \citet{DelgadoMena19}, who found that the influence of [Fe/H] abundances on age-[X/Fe] relations varies strongly per element, such that (for example) the age-[Mg/Fe] relation is relatively consistent across different metallicities, while the relations for $s$-process elements such as [Y/Fe] are not. Thus, while the influence of varying stellar metallicities must be accommodated for, the results of \citet{DelgadoMena19} suggest that chemical clocks remain a valid method for stellar age estimation. We therefore aim to use chemical clocks to constrain the age of the \thisstar{} system.

The general process for calibrating chemical clocks is to collect abundance data for a sample of stars whose isochronal ages are well-constrained (often stars physically similar to the Sun, as in e.g. \citealt{Nissen15, TucciMaia16}, as here stellar models are typically most precise), and then fitting a relation between isochronal ages and [X/Fe] abundance ratios for selected elements. In this work we make use of the abundances and ages from \citet{Brewer16}, based on observations of a large sample of stars with the HIRES spectrograph, as this study is one of the few to provide abundance measurements for \thisstarAa{}. Previous studies on abundance-age trends were based on data from other instruments (e.g. HARPS in \citealt{Nissen15}, \citealt{DelgadoMena19}), and as amply demonstrated by \citet{Brewer16} there are systematic differences in abundance measurements as measured by different spectrographs. It is therefore not possible to directly compare the measured abundances of \thisstarAa{} to the chemical clock relations from previous studies based on other data; we must instead measure the age-abundance trends for stars in the \citet{Brewer16} sample \textit{de novo}.

We apply a series of cuts to the \citet{Brewer16} sample to select stars with precise ages and similar physical parameters to \thisstar{} (particularly metallicity). We detail the selection in Appendix~\ref{appendix:chemical_clocks}. A total of 73 stars survive our cuts, and for this sample we inspect the age-[X/Fe] plots for the elemental abundances measured by \citet{Brewer16} to search for relationships that would be useful as chemical clocks. Among available abundance measurements, we observe clear age trends in the $\alpha$-elements [Mg/Fe] and [Al/Fe], as has been found ubiquitously in previous studies on chemical clocks. However, we surprisingly do not observe a significant age-[Y/Fe] relation, unlike previous studies such as \citet{Nissen15}. If there is such a correlation in our sample, its amplitude is smaller than the scatter of the data ($\sigma_{\text{[Y/Fe]}}=0.09$). As a result of this, an age-[Y/Mg] relation for our sample performs worse than an age-[Mg/Fe] alone.

Additionally, there is good reason to exclude consideration of [Y/Fe] for our purposes. \citet{TucciMaia16} and \citet{Spina18} both find that the star HIP~64150 (= HD~114174) is anomalously rich in $s$-process neutron-capture elements such as yttrium, and is thus a marked outlier in the age-[Y/Fe] relation in both studies. The source of these anomalies is undoubtedly the white dwarf companion HD~114174~B \citep{Crepp13, Gratton21}, as wind accretion during the asymptotic giant branch stage of evolution of the white dwarf progenitor \citep{Boffin88} can explain the overabundance of $s$-process elements for stars with white dwarf companions \citep{Fuhrmann19}. As our target system likewise contains a white dwarf component, there is reason to assume that the age-[Y/Fe] relation would be invalid for \thisstarAa{} even if there was a detectable trend in the abundance data.

We therefore disregard [Y/Fe] and focus on the [Mg/Fe] and [Al/Fe] chemical clocks. Of the 73 stars in our sample, all have Fe and Mg abundance measurements in \citet{Brewer16} whereas three stars (HD~49933, HD~82328, HD~168151) lack Al abundances. We choose to exclude the Al measurements for a further three stars (HD~120064, HD~209253, HD~199260) as they have anomalously low abundance ratios ([Al/Fe] < -0.3, making them >3$\sigma$ outliers) that tend to skew the fitted age-abundance relationship. The age and abundance ratio data for our sample can be found in Appendix~\ref{appendix:chemical_clocks}. As for the abundance uncertainties, \citet{Brewer16} estimate statistical uncertainties of $\sigma$[Fe/H] $=0.010$, $\sigma$[Mg/H] $=0.012$, and $\sigma$[Al/H] $=0.028$. Extending these values to the abundance ratios under the assumption that the elemental uncertainties are uncorrelated and strictly Gaussian, we calculate $\sigma=0.016$ for [Mg/Fe] and $\sigma=0.030$ for [Al/Fe] respectively.

We fit the age-abundance relationship using a simple linear relationship:

\begin{equation}
    \text{[X/Fe]} = a+b(\text{Age})
    \:,
\end{equation}

Where $a$ and $b$ are coefficients specific to the particular abundance ratio [X/Fe] under consideration.

\begin{figure*}
	\includegraphics[width=\columnwidth]{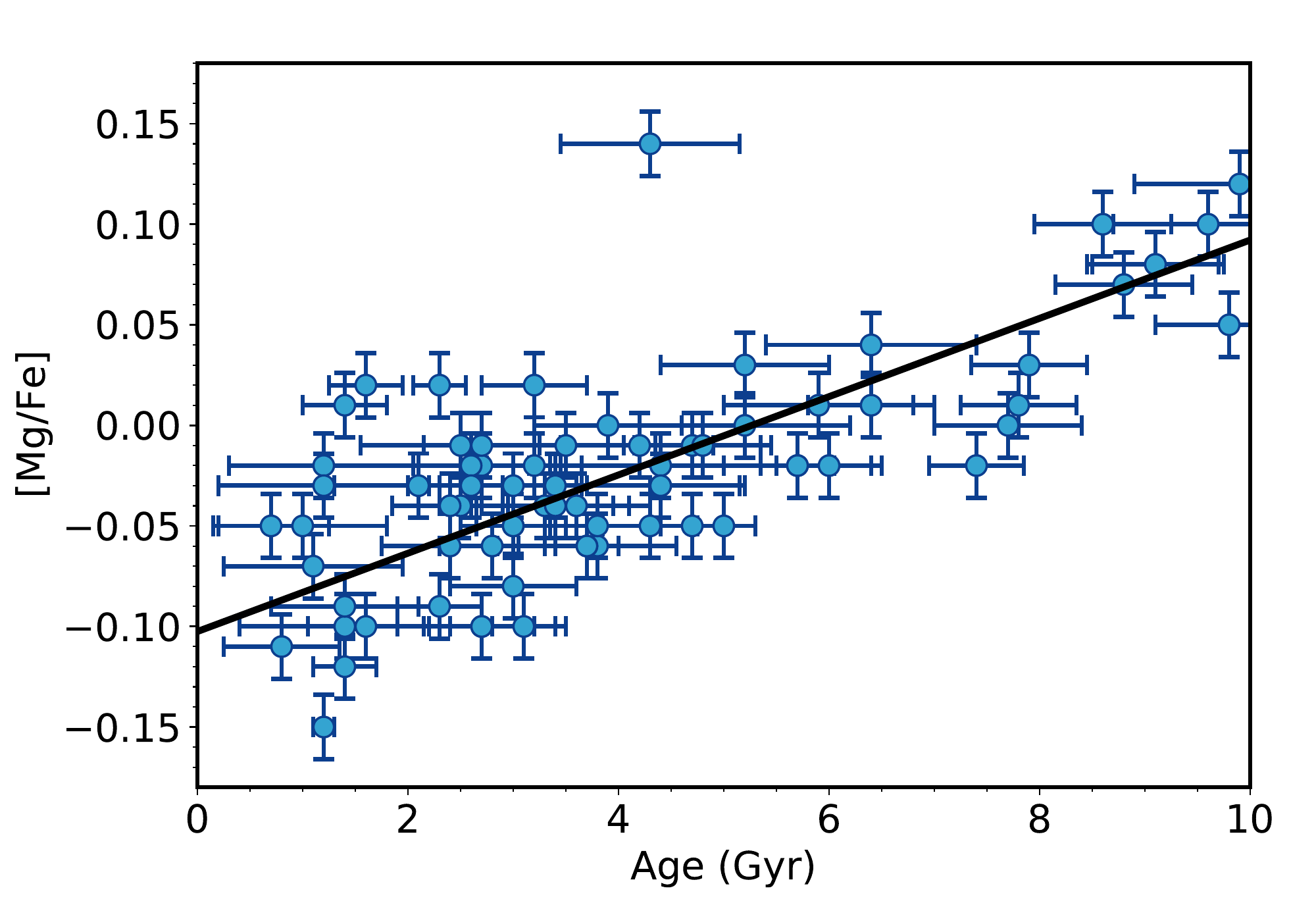}
	\includegraphics[width=\columnwidth]{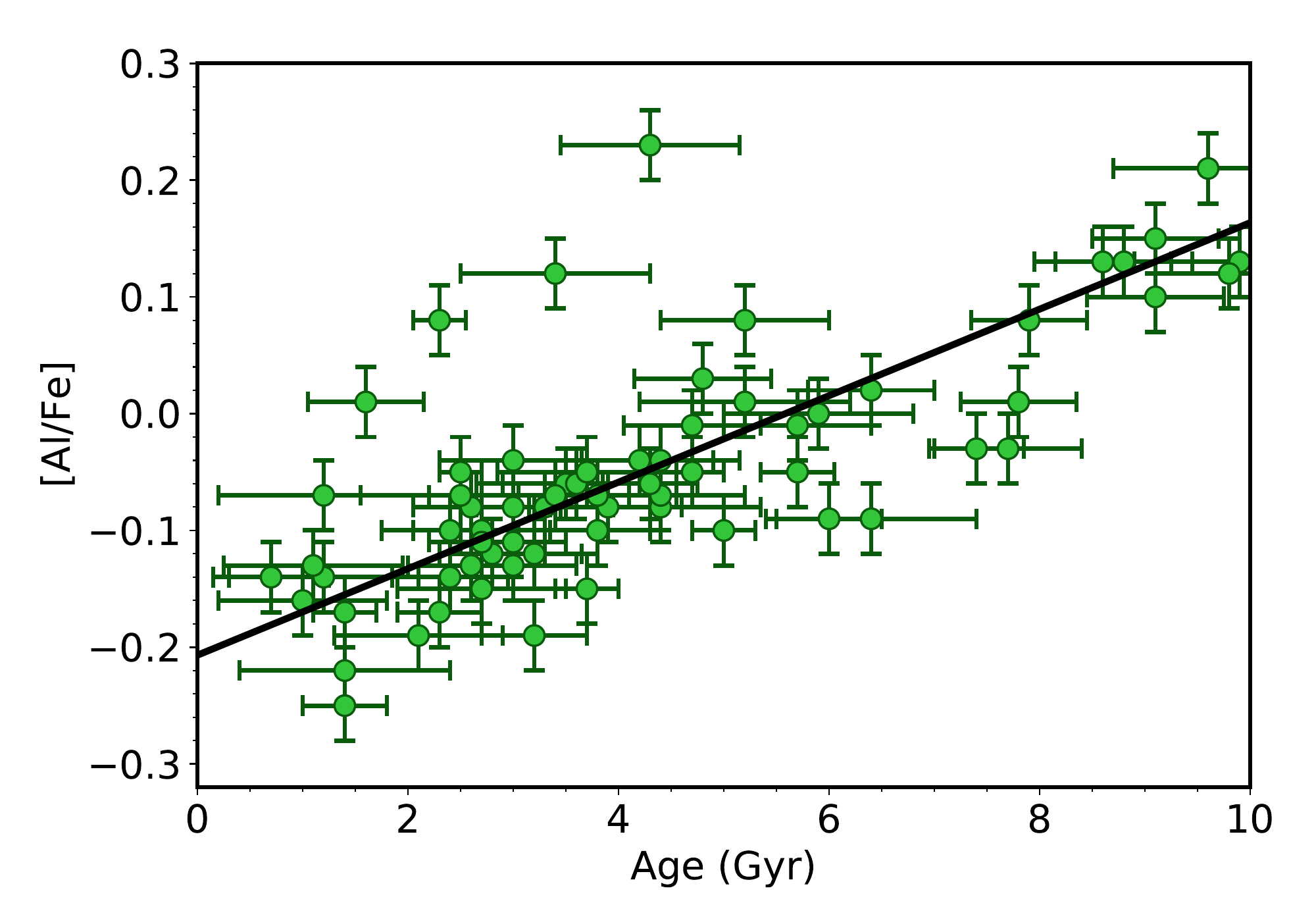}
	\caption{Age-abundance relationships in [Mg/Fe] (left) and [Al/Fe] (right) for our sample of stars selected from \citet{Brewer16}.}
	\label{figure:abundance_trends}
\end{figure*}

\begin{table*}
	\centering
	\caption{Parameters of our fitted age-abundance relationships.}
	\label{table:abundance_trends}
	\begin{tabular}{lcccc}
		\hline
		Abundance & $a$ (dex) & $b$ (dex Gyr$^{-1}$) & $\sigma$[X/Fe] (dex) & $\sigma$Age (Gyr) \\
		\hline
		$\text{[Mg/Fe]}$ & $-0.103\pm0.006$ & $0.0195\pm0.0013$ & $0.039$ & $2.0$ \\
		$\text{[Al/Fe]}$ & $-0.209\pm0.011$ & $0.0374\pm0.0024$ & $0.071$ & $1.9$  \\
		\hline
	\end{tabular}
\end{table*}

We tabulate the parameters of our fits in Table~\ref{table:abundance_trends} and show the best-fit age-abundance relationships in Figure~\ref{figure:abundance_trends}. We obtain robust correlations between abundance ratio and age for both [Mg/Fe] and [Al/Fe]. Though the uncertainties and scatter for [Al/Fe] are approximately twice as large as for [Mg/Fe], the slope of the age-abundance correlation is also twice as large, hence the S/N ratio for the two abundance trends are approximately equal.

Having now established our chemical clocks, the next step is to invert the relationships to estimate the age of \thisstar{}. \citet{Brewer16} measure elemental abundances of $\text{[Fe/H]}=-0.25$, $\text{[Mg/H]}=-0.21$, $\text{[Al/H]}=-0.15$ for \thisstarAa{}, hence we assume stellar abundance ratios of $\text{[Mg/Fe]}=0.04$ and $\text{[Al/Fe]}=0.10$. As \citet{Brewer16} do not provide star-specific estimates for abundance uncertainties, it would appear justifiable to assume the statistical uncertainties estimated for the chemical clock sample, i.e. $\sigma_{\text{[Mg/Fe]}}=0.016,\:\sigma_{\text{[Al/Fe]}}=0.030$. This can be supported with the low values of scatter for the spectral fit reported by those authors for \thisstarAa{} ($\text{C-rms}=0.01$ and $\text{L-rms}=0.01$). However, their spectral fit does not account for the flux contribution from \thisstar{}~Ab and B, which \citet{Fuhrmann19} found to have a small but significant effect on the measurement of [Fe/H] in their observations. The impact of the companion flux contribution on the abundance measurements of \citet{Brewer16} cannot be quantified without further study, but in recognition of the possible inaccuracy of the abundances we opt to double the uncertainties on the abundance ratios. This results in abundance ratios of $\text{[Mg/Fe]}=0.04\pm0.032$ and $\text{[Al/Fe]}=0.10\pm0.06$ for \thisstarAa{}.

Feeding these abundance ratios into our age-abundance relationships, we derive chemical clock ages for \thisstar{} of $7.3\pm1.7$~Gyr from [Mg/Fe] and $8.3\pm1.6$~Gyr from [Al/Fe]. However these estimates are likely to have underestimated uncertainties, as this calculation does not adequately account for the scatter in our age-abundance relations. We therefore add in the standard deviations in age of our fits given in Table~\ref{table:abundance_trends} in our age estimation to represent the scatter of our fits. This results in final chemical clock ages for \thisstar{} of $7.3\pm2.6$~Gyr and $8.3\pm2.5$~Gyr based on the [Mg/Fe] and [Al/Fe] abundances respectively.

\subsubsection{Activity age} \label{subsec:age_activity}

Sun-like stars are thought to have magnetic fields similar in nature to the field presented by the Sun, i.e. driven by a stellar dynamo which is connected to its rotation \citep{Charbonneau13}. As the star ages it is expected to lose angular momentum as a result of wind-driven mass loss \citep{Weber67}, in turn leading to weakening of the stellar magnetism. Stellar age, rotation, and magnetic activity are therefore thought to be interconnected for Sun-like stars; and as the latter two items are (indirectly) observable, this tripartite relationship provides a valuable means for estimating the ages of stars - particularly Sun-like main sequence stars - whose ages are otherwise difficult to determine.

Study of the age-rotation-activity connection began decades ago \citep{Kraft67, Skumanich72} and continues down to the present. In practice the age-rotation relation and age-activity relation are often studied independently (e.g. \citealt{Barnes03, Delorme11, Meibom15, Curtis20} for age-rotation, \citealt{Soderblom91, LorenzoOliveira16, LorenzoOliveira18} for age-activity), however some studies have considered the connection between all three terms \citep{Mamajek08b, Wright11}.

In this study we focus on the age-activity relation for age estimation, as the rotational period of \thisstarAa{} has not presently been measured. One of the main parameters used to quantify stellar activity in the literature is the $\log R^{\prime}_{HK}$ index, which is defined based on the strength of the calcium H \& K emission lines \citep{Wright04}. Measurements of this activity indicator for \thisstarAa{} in the literature have consistently found low values ($\log R^{\prime}_{HK}=-4.96$, \citealt{Wright04}; $\log R^{\prime}_{HK}=-4.85$, \citealt{Isaacson10}), suggesting a relatively old age for the system. Most recently, \citet{StanfordMoore20} have calibrated the $\log R^{\prime}_{HK}$-age relationship for Sun-like stars based on activity measurements of members of open clusters with precisely constrained ages, and used their results to estimate the ages of field stars. For \thisstarAa{}, \citet{StanfordMoore20} use their age-activity relation to estimate a stellar age of $6.4^{+3.2}_{-2.6}$~Gyr assuming a $\log R^{\prime}_{HK}$ value of -4.875, which agrees well our age estimates derived using other methods.

However, as was the case for the [Y/Mg] chemical clock, the structure of the \thisstar{} system allows for doubt that magnetic activity is a reliable age indicator in this instance. It was suggested as early as \citet{Wegner73} that mass loss of a white dwarf progenitor may cause the revitalisation of magnetic activity of a nearby stellar companion, a hypothesis which can be demonstrated using a nearby Sirius-like system as an example. \citet{Zurlo13} presented the discovery of a white dwarf companion to the K2V star HD~8049, found as part of a targeted search for substellar companions to young stars with direct imaging \citep{Chauvin15}. HD~8049 was identified as a young star due to rapid rotation and high stellar activity; however, \citet{Zurlo13} observe a low stellar lithium abundance inconsistent with a young star and find that the kinematics of HD~8049 are instead similar to stars with ages of a few Gyr. The authors interpret these discordant age indicators as a result of a spin-up of HD~8049~A due to mass accretion from the progenitor of the white dwarf companion, causing the K-dwarf to appear rejuvenated in its magnetic activity. Indeed, rotational spin-up appears to be a ubiquitous phenomenon for stars with nearby white dwarf companions \citep[][]{Fuhrmann14}, to the extent that \citet{Leiner18} observed that the spin-down of such post-accretion stars is quantitively similar to spin-down after star formation based on a sample of 12 post-mass-transfer binaries, making it appear that the post-accretion star has been magnetically and rotationally reborn.

The aforementioned post-mass-transfer systems are typically more tightly spaced than \thisstar{}, where the A-C separation is at least $\geq$590~AU. \boldnew{Indeed, to our knowledge the limiting separation for which stellar activity can be significantly increased due to wind accretion has not previously been explored.} Nevertheless, we consider it possible that \thisstarAa{} absorbed some fraction of its mass from the giant progenitor of \thisstarC{} via wind accretion, which could then have revitalised the magnetic activity of the primary star. This would make the $\log R^{\prime}_{HK}$ of \thisstarAa{} invalid as an age indicator, and we therefore advocate caution in the interpretation of the activity age for this system.

\subsection{Summary of age constraints} \label{subsec:age_summary}

\begin{table*}
	\centering
	\caption{Summary of our adopted age estimates for \thisstar{}. The first five rows reflect the age estimates from Section~\ref{subsec:system_age}, the next two are averages of the kinematic and chemical clock age estimates, and the final row is the product of the isochronal, kinematic, and chemical clock ages.}
	\label{table:age_summary}
	\begin{tabular}{lccccc}
		\hline
		Method & \multicolumn{5}{c}{Age CDF posterior value (Gyr)} \\
		& 2.5\% (-2$\sigma$) & 16\% (-1$\sigma$) & 50\% & 84\% (+1$\sigma$) & 97.5\% (+2$\sigma$) \\
		\hline
		Isochrone fit & 0.8 & 2.6 & 5.6 & 8.9 & 12.0 \\
		Kinematics (\citetalias{AlmeidaFernandes18} relation) & 2.4 & 4.4 & 7.8 & 11.7 & 13.6 \\
		Kinematics (\citetalias{Veyette18} relation) & 4.6 & 6.8 & 9.6 & 12.3 & 13.7 \\
		$\text{[Mg/Fe]}$ chemical clock & 2.2 & 4.7 & 7.3 & 9.9 & 12.4 \\
		$\text{[Al/Fe]}$ chemical clock & 3.4 & 5.8 & 8.3 & 10.8 & 13.2 \\
		Activity \citep{StanfordMoore20} & 1.8 & 3.7 & 6.4 & 9.6 & 12.3 \\
		\hline
		Kinematic average $^a$ & 2.8 & 5.2 & 8.7 & 12.0 & 13.6 \\
		Chemical clock average $^b$ & 2.8 & 5.3 & 7.8 & 10.3 & 12.8 \\
		\hline
		Product of age probability distributions $^c$ & 3.8 & 5.5 & 7.3 & 9.2 & 11.0 \\
		\hline
		\multicolumn{6}{l}{$^a$ Mean of the two kinematic age estimates. $^b$ Mean of the two chemical clock age estimates. $^c$ Product of the} \\
		\multicolumn{6}{l}{probability distributions for the isochronal age, the average kinematic age, and the average chemical clock age.} \\
	\end{tabular}
\end{table*}

\begin{figure}
    \vspace{6mm}
	\includegraphics[width=\columnwidth]{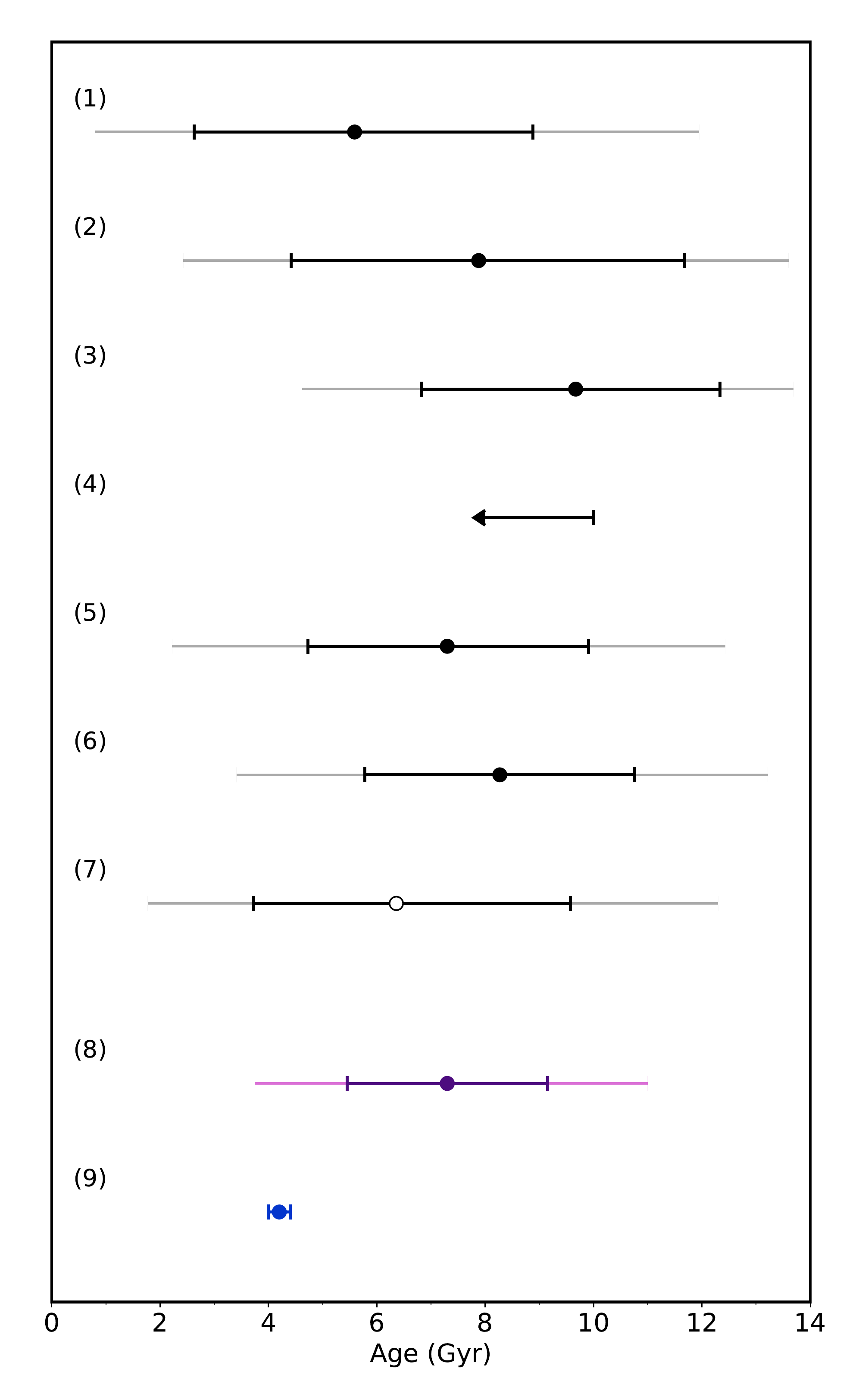}
	\caption{Graphical summary of our adopted age constraints for \thisstar{}. From top to bottom, these ages are derived from (1) isochrone fitting; (2) the \citetalias{AlmeidaFernandes18} $UVW$-age relation; (3) the \citetalias{Veyette18} $UVW$-age relation; (4) an assumed upper limit for the age of the thin disk; (5) the [Mg/Fe] chemical clock; (6) the [Al/Fe] chemical clock; and (7) the stellar activity-age relationship of \citet{StanfordMoore20}. Points and bars in black show the medians and 1$\sigma$ uncertainties of the age estimates, while the grey bars reflect the 2$\sigma$ uncertainties. The last of these is highlighted in white since we consider it possible that the activity age is invalid for this system (see Section~\ref{subsec:age_activity}), and this age is not used in further calculations. Under (8) is the $7.3^{+1.9}_{-1.8}$~Gyr age arising from the product of the foregoing age estimates for the system, while the point under (9) is our $4.2\pm0.2$~Gyr estimate for the total age of the white dwarf \thisstarC{} not accounting for \Ne{} distillation. While the difference between these values is not statistically significant ($p=0.05,\:1.7\sigma$), it is also not inconsistent with the theoretical expectation of a $\approx$1~Gyr white dwarf cooling anomaly caused by \Ne{} distillation.}
    \vspace{6mm}
	\label{figure:age_summary}
\end{figure}

In the preceding sections we have employed a variety of techniques to constrain the age of the \thisstar{} system for the purpose of attempting to detect an anomaly in the cooling age of the crystallising white dwarf \thisstarC{}. We summarise our results in Table~\ref{table:age_summary} and Figure~\ref{figure:age_summary} by showing the medians, 1$\sigma$ confidence intervals, and 2$\sigma$ confidence intervals for the cumulative distribution function (CDF) of the age probability distribution for each method of age estimation (except for the constraint from the age of the thin disk, which we parameterise as an upper limit of $<$10~Gyr). The uncertainties for all of our system age estimates are relatively large (median 1$\sigma$ uncertainty~=~2.8~Gyr) but it can be seen that they are generally consistent with each other. All of the medians for our system ages are larger than the total age of $4.2\pm0.2$~Gyr for \thisstarC{} calculated in Section~\ref{subsec:WD_parameters}, however only for the \citetalias{Veyette18} kinematic age can this be placed beyond the 95\% confidence interval.

To improve our resolution on the system age we next aim to combine the constraints derived from different techniques. For this purpose we opt to exclude the activity age, since as discussed in Section~\ref{subsec:age_activity} there is a possibility that this age indicator may be invalid for the \thisstar{} system. We further exclude the $<$10~Gyr upper limit from the age of the thin disk as this would render our age posterior non-Gaussian, leading to a bias towards younger ages. As a first step we calculate the mean of the two kinematic ages and the two chemical clock ages to avoid over-weighting age estimates which are not truly independent; this results in average kinematic and chemical clock ages of $8.7^{+3.3}_{-3.5}$~Gyr and $7.8\pm2.5$~Gyr respectively. Taking the product of the probability distributions for the isochronal, kinematic, and chemical clock ages, we derive a combined age of $7.3^{+1.9}_{-1.8}$~Gyr (1$\sigma$) which we adopt as our final age estimate for the \thisstar{} system.\footnote{Including the $<$10~Gyr upper limit on the age results in a slightly lower estimate of $7.1^{+1.6}_{-1.7}$~Gyr.} Compared to the $4.2\pm0.2$~Gyr total age of \thisstarC{} neglecting \Ne{} distillation, we calculate an age anomaly of $+3.1\pm1.9$~Gyr between the two estimates. Due to the large uncertainty on the system age the age difference is not statistically significant ($p=0.05,\:1.7\sigma$), so we cannot yet claim to have empirically detected an anomaly in the cooling of \thisstarC{}.

\section{Discussion} \label{sec:discussion}

\subsection{A new Sirius-like system in the solar neighbourhood}

In this work we have reported the discovery of a new Sirius-like system at 32~parsecs, \thisstar{}, composed of a compact main sequence triple plus a widely separated ($\geq$590~AU) white dwarf. \citet{Holberg13}, who provided the most recent review of Sirius-like systems, were aware of only 21 SLSs within 40 parsecs of the Sun; while that number has been increased in subsequent years thanks to discoveries from adaptive optics imaging \citep[e.g.][]{Hirsch19, Bonavita20, Bowler21} and from \textit{Gaia} data as in this study \citep{Scholz18, Landstreet20}, Sirius-like systems remain relatively uncommon and the addition of a new member to this sample is significant. In the sample of SLSs known to \citet{Holberg13} there was a significant decline in the frequency of systems beyond 20 parsecs, suggesting low completeness at larger distances. Remarkably, all of the aforementioned Sirius-like systems discovered subsequent to that work lie within 20-40~pc -- including \thisstar{} -- suggesting that this gap is being filled out by new discoveries.

As a quadruple system, \thisstar{} has among the highest-order multiplicities of any known SLS. In the \citet{Holberg13} sample the only quadruple Sirius-like system is 56~Persei, whereas 14~Aurigae is quintuple \citep[both systems from][]{Barstow01}. The nearby SLS HD~6101 (missed in \citealt{Holberg13}) is also a quadruple system, composed of a K-type visual binary in an extremely wide pair (1280" separation, $\geq$27000~AU projected) with the double white dwarf WD~0101+048 \citep{Maxted00, Makarov08}. Finally, though it was not recognised in \citet{Crepp18}, the recently discovered Sirius-like binary HD~169889 forms a wide common proper motion pair with HD~169822 \citep[610" separation, $\geq$21600~AU;][]{Shaya11}, which is in turn a spectroscopic binary with a 293-day orbital period \citep{Vogt02, Halbwachs12}, making this a quadruple system as well. Thus, with only four Sirius-like systems with quadruple or higher multiplicity being known to us, the addition of the \thisstar{} system to this small class is notable.

\subsection{HD 190412 C, a benchmark crystallising white dwarf} \label{subsec:WD_discussion}

In Section~\ref{subsec:WD_parameters} we presented an analysis of the fundamental parameters of \thisstarC{}. We found that the white dwarf has an effective temperature of $T_{\text{eff}}=6600\pm80$~K and a mass of $M=0.817\pm0.019~M_\odot$, which places it in an area of the temperature-mass plane predicted to be occupied by white dwarfs undergoing core crystallisation (assuming C/O core composition; see Figure~\ref{figure:WD_fit}). This makes \thisstarC{} the first crystallising white dwarf \boldnew{belonging} to a Sirius-like system to be \boldagain{confirmed} as such. Furthermore, the star lies in the pile-up of white dwarfs with a core crystallisation fraction of $\approx$60\%, a feature of the white dwarf population \boldnew{that is clearly identifiable in \citet[figures 18, 19]{Kilic20}} and argued to be a result of \Ne{} phase separation \boldnew{by \citet{Blouin21}}. These facts establish \thisstarC{} as an important benchmark for understanding white dwarf crystallisation.

The association of \thisstarC{} with the main sequence stars \thisstarAB{} makes this the first identified crystallising white dwarf whose total age can be externally constrained,\footnote{\boldagain{It has been pointed out to us by the anonymous reviewer that \citet{Heintz22}, in their study of age estimates of WD+WD binaries, note the likely presence of crystallised white dwarfs in their sample. While this was not studied in detail, we do not dispute that it is likely many of their white dwarfs are undergoing crystallisation. However, it is not possible to constrain the \textit{total} ages of these systems without reference to the IFMR for both white dwarfs, which results in a double model dependence that makes it infeasible to use these systems to measure cooling delays as pursued in this work.}} meaning that it is possible in principle to empirically detect a delay in its cooling by comparing the model age of the white dwarf against the age of the system. By combining its white dwarf cooling age with the theoretical lifetime of its progenitor we estimated a total age of $4.2\pm0.2$~Gyr for \thisstarC{} in Section~\ref{subsec:WD_lifetime}. We then applied several different techniques for age estimation to \thisstarAB{} (isochrones, kinematics, chemical clocks, and stellar activity), resulting in a final system age of $7.3^{+1.9}_{-1.8}$~Gyr. Comparison of these two age estimates results in an empirical age anomaly of $+3.1\pm1.9$~Gyr; this is not statistically significant in its own right ($p=0.05$), but is consistent with expectations of a cooling delay for \thisstarC{}. We hereon refer to the age anomaly as a "tension", by which we intend to recognise that our age estimates for the white dwarf and the system are not \boldagain{statistically} incompatible, yet are suggestive of underestimation of the age of \thisstarC{}.

Before exploring the possibility that the age tension is consistent with the effects of \Ne{} distillation, we first consider sources of uncertainty in the estimation of the white dwarf age. Many of these can be dismissed:

\begin{enumerate}
\item The \boldnew{lifetime of a main sequence star} is related to its mass by an approximate negative exponential function. As a result, uncertainties in the progenitor lifetime can dominate the total age uncertainty for lower-mass white dwarfs with near-solar progenitor masses \citep[e.g.,][]{Heintz22}. However, \thisstarC{} is a relatively massive white dwarf ($0.817\pm0.019~M_\odot$), which in turn implies a super-solar-mass progenitor ($3.38^{+0.13}_{-0.10}~M_\odot$) with a short pre-white dwarf lifetime ($0.29\pm0.03$\,Gyr). The uncertainty on the progenitor lifetime therefore contributes only a small amount to the total age uncertainty.

\item Until now, we have assumed single-star evolution for \thisstarC{}. However, it is important to consider the possibility that it has experienced a stellar merger at some stage of its evolution. \boldnew{White dwarfs resulting from mergers prior to the white dwarf stage (i.e. before the end of nuclear fusion, in contrast to $\text{WD}+\text{WD}$ mergers) are thought to exist \citep[e.g.][]{Andrews16} The simulations of \boldagain{\citet{Temmink20}} suggest that as many as $\approx$20\% of white dwarfs may originate from mergers prior to the white dwarf stage. If it can be assumed that a white dwarf that forms from such a post-merger star obeys the same IFMR as normal WDs, it could be posited in the extreme case that \thisstarC{} ($M_{\text{init}}=3.4~M_\odot$) formed from the merger product of two $1.7+1.7~M_\odot$ stars. If this were the case, then the pre-white dwarf lifetime of \thisstarC{} could be extended to as much as $\sim$3~Gyr. As there does not appear to be any way to tell if an individual white dwarf results from this kind of merger, it cannot be excluded that \thisstarC{} results from a pre-WD merger product. However, this scenario is certainly less probable than single-star evolution \textit{a priori}.}

\item \boldnew{Though $\text{WD}+\text{WD}$ mergers are likely to be rarer than pre-WD mergers (\citealt{Temmink20} estimate that only $\sim$3\% of white dwarfs form in this manner), they could be more pernicious\boldagain{, since} $\text{WD}+\text{WD}$ mergers can occur at any system age. There is substantial observational evidence suggesting that a large fraction of} ultra-massive white dwarfs \boldnew{are} $\text{WD}+\text{WD}$ merger products \citep{Kilic21}\boldagain{. However}, \citet{Cheng20} estimate that only $\approx$10\% of 0.8-0.9~$M_\odot$ white dwarfs have formed via \boldnew{$\text{WD}+\text{WD}$} mergers, suggesting that such an origin for \thisstarC{} is unlikely \textit{a priori}. Furthermore, post-merger white dwarfs often display peculiar properties, such as unusual chemical compositions \citep{Coutu19, Hollands20}, strong magnetic fields \citep{GarciaBerro12, Caiazzo21}, and rapid rotation \citep{Pshirkov20, Reding20, Caiazzo21, Kilic21.rotation}. Though we are unable to measure its rotational period with the data available, no peculiarities of the atmospheric composition or magnetism of \thisstarC{} are evident to us. While the absence of evidence for a merger cannot be considered proof that the star has \textit{not} experienced such a phase, we conclude that \boldnew{there is no reason to believe that \thisstarC{} results from a $\text{WD}+\text{WD}$ merger}.

\item Ordinarily the metallicity of a white dwarf is inaccessible, which results in uncertainty on the abundance of $X(\Ne{})$ in the core. However, for a white dwarf in a Sirius-like system the measurable metallicity of the non-degenerate component can be assumed to represent the primordial metallicity of the white dwarf. \thisstarAa{} has a relative metal abundance of $\text{[M/H]}=-0.24$ \citep{Brewer16}, which we have used to inform our assumed \Ne{} mass fraction for \thisstarC{} ($X(\Ne{})=0.008$, Section~\ref{subsec:cooling_model}). This eliminates the possibility that we are underestimating the white dwarf cooling age because of an underestimation of its \Ne{} content (which would in turn imply an underestimation of the cooling delay associated with \Ne{} settling in the liquid phase).

\item Current uncertainties on the thermal conductivity of the envelope also have a small effect on the inferred age of \thisstarC{}. The cooling model used to determine the age of \thisstarC{} relies on the conductive opacities of \cite{Cassisi07}, with the corrections of \cite{Blouin20b} in the partially degenerate regime. \cite{Cassisi21} have pointed out that there are different plausible prescriptions to bridge the results of \cite{Blouin20b} with the standard opacities of \cite{Cassisi07} in the strong degeneracy limit. This introduces uncertainties on white dwarf cooling ages that can reach 25\% at low luminosities. The maximum age uncertainty due to this bridging problem can be obtained by comparing cooling tracks that include the \cite{Blouin20b} corrections to cooling tracks that directly use the \cite{Cassisi07} conductivities without any corrections. For \thisstarC{}, we find that ignoring the \cite{Blouin20b} corrections only leads to a 0.4\,Gyr increase of the cooling age.

\item Because of uncertainties on the $^{12}{\rm C}(\alpha,\gamma)^{16}{\rm O}$ reaction rate \citep{deBoer17} and, even more importantly, on the efficiency of convective boundary mixing during core helium burning \citep{Straniero03,Salaris10,Constantino15,Giammichele18,DeGeronimo19,Ostrowski21}, the $X({\rm O})$ profile of white dwarf cores remains a poorly constrained quantity. This uncertainty on $X({\rm O})$ induces an additional uncertainty on white dwarf cooling ages as it affects the thermal content of the white dwarf and the energy released by C/O phase separation during crystallisation. To verify whether this uncertainty could explain the age tension, we computed additional cooling tracks assuming different $X({\rm O})$ values for the uniform core composition of  \thisstarC{}. Varying $X({\rm O})$ within a reasonable range of parameters ($0.5 \leq X({\rm O}) \leq 0.8$), we found that the cooling age of \thisstarC{} never changed by more than 0.2\,Gyr. Assuming a more realistic non-uniform $X({\rm O})$ profile could affect the age more significantly. However, this could only lead to a decrease of the age since a uniform profile necessarily maximizes the effect of C/O phase separation \citep[e.g.,][]{Salaris09}. We can therefore conclude that any uncertainty on the $X({\rm O})$ profile cannot explain the age tension.

\item The thickness of the superficial H and He layers is also a well-known source of uncertainty for white dwarfs age dating. Our fiducial model of \thisstarC{} assumed a standard $10^{-2}\,M_{\star}$ He envelope and a thick $10^{-4}\,M_{\star}$ H layer. Varying those values within a reasonable range of parameters for a DA white dwarf of this temperature ($10^{-2}-10^{-3}\,M_{\star}$ for the He envelope and $10^{-4}-10^{-8}\,M_{\star}$ for the H layer, \citealt{Rolland18}), we find that the cooling age of \thisstarC{} never changes by more than 0.3\,Gyr.
\end{enumerate}

Having \boldnew{determined} that \boldnew{it is improbable or infeasible for} these known sources of uncertainty \boldnew{to} significantly influence the age estimation \boldnew{of} \thisstarC{}, we now examine whether \Ne{} distillation, so far omitted from our cooling models, could explain the age tension. Precisely determining the cooling age of \thisstarC{} including the effect of \Ne{} distillation would require better constraints on the $X({\rm O})$ profile of its core. The exact $X({\rm O})$ profile will determine when \Ne{} distillation takes place (see figure~3 of \citealt{Blouin21}), which will in turn affect the resulting cooling delay for at least three reasons: (1) the earlier it takes place, the higher the luminosity $L$ of the star, which will tend to decrease the cooling delay ($\Delta \tau \sim \Delta B / L$, where $\Delta B$ is the change of binding energy due to distillation); (2) the earlier it takes place, the larger the quantity of \Ne{} available for distillation in the liquid layers of the core (the \Ne{} in the frozen layers is not available for distillation), which will increase $\Delta B$; (3) the distillation process and its accompanying cooling delay may still lie ahead in the evolution of \thisstarC{}, be underway, or already be completed. Nevertheless, we can still reasonably evaluate the additional cooling delay from \Ne{} distillation using the estimates of \citet[table~1]{Blouin21} and adjusting for the \Ne{} mass fraction of \thisstarC{}. We find that \Ne{} distillation could add an additional cooling delay of up to $\approx$1~Gyr to the age of \thisstarC{}. This is substantially larger than the previously discussed sources of age uncertainty and is fully consistent with our $+3.1\pm1.9$~Gyr empirical age anomaly. Thus, while the tension between our age estimates is not significant enough to require additional mechanisms of energy release during crystallisation in its own right, we conclude that the \Ne{} distillation cooling delay hypothesis provides a consistent and physically plausible mechanism for why the age of \thisstarC{} may be underestimated by conventional models.

\subsection{Prospects for future studies}

In this work we have presented the discovery and analysis of the first Sirius-like system containing a crystallising white dwarf to be \boldagain{confirmed} in the literature. We argue that white dwarfs in such systems are uniquely valuable calibrators of crystallisation models by virtue of the fact that they form the only substantial population of local white dwarfs in the appropriate mass and temperature range whose total ages can be externally constrained \citep[with the exception of the low-mass white dwarfs in old clusters, for which the effects of crystallisation are difficult to disentangle from contemporaneous convective coupling;][]{Bergeron19, Tremblay19}. We have therefore made a concerted attempt to constrain the age of the \thisstar{} system in order to test white dwarf cooling models. However, as discussed in the preceding section, our final estimate for the system age is insufficiently precise ($7.3^{+1.9}_{-1.8}$~Gyr) to detect a statistically significant anomaly in the white dwarf cooling age. We suggest that accomplishing this would require an age uncertainty below $\lessapprox$1~Gyr for this system.

It is certainly possible that further study will allow for improvement on the precision of the age of \thisstar{} over the levels that we have achieved. While significant improvement on the kinematic age precision is unlikely due to the inherently statistical nature of the method \citep{AlmeidaFernandes18}, we believe that more precise isochronal and chemical clock age estimates are possible. The uncertainties on the chemical clock ages primarily reflects the uncertainties on the abundances and the isochronal ages for the calibrator stars, which could be improved by increasing the sample size and acquiring higher abundance precisions with higher S/N spectra. Additionally, for \thisstar{} in particular, a model of the stellar spectrum accounting for the flux contributions from the faint companion stars would help to reduce any systematics present in the abundance measurements. For the isochronal age the main sources of uncertainty is the mass of \thisstarAa{}, as made evident by the strong anti-correlation between these parameters in Figure~\ref{figure:age_isochrone}. The architecture of the system offers a distinct advantage in this respect, since it is possible to dynamically constrain the masses of the stars in the inner triple. If this can be used to reduce the mass uncertainty for \thisstarAa{} significantly below our isochrone-only value ($M=0.87\pm0.03~M_\odot$), it could theoretically be used to more precisely estimate the age of the system. However at this point it would become necessary to account for systematic differences between isochrone models, which result in age uncertainties on the order of $\approx$20\% for main-sequence stars \citep{Tayar22}. Correctly accounting for these systematic uncertainties will necessarily result in a noise floor for the isochronal age. In summary, while improvements in the age precision for \thisstar{} are certainly possible, it is difficult to envision this resulting in an age uncertainty below $\lessapprox$1~Gyr as might be required to detect a statistically significant white dwarf cooling anomaly.

We therefore point to discovery of similar systems as a promising avenue for expanding on our results. \thisstar{} lies at a distance of only 32~parsecs from the Sun, and it is undoubtedly likely that other Sirius-like systems containing crystallising white dwarfs remain be discovered in the solar neighbourhood. If we assume that the space density of such systems is approximately 1 per 32~$\text{pc}^3$, then there should be $\approx$30 within 100~pc of the Sun. Crystallising white dwarfs can be easily identified from \textit{Gaia} photometric data \citep{Tremblay19} and gravitationally bound Sirius-like systems can likewise be discovered using \textit{Gaia} astrometry, so a targeted search for these systems in the \textit{Gaia} catalogue would be an efficient method for their discovery. It can be expected that this sample will contain stars more amenable to age-dating than \thisstar{} (e.g. evolved stars which are better suited for isochronal age estimation than dwarfs), or stars whose ages can be constrained using techniques other than those used in this work \citep[e.g. asteroseismology;][]{Aerts15}, meaning that for those systems it would be more feasible to detect anomalies in white dwarf cooling ages. Moreover, assembling a larger sample of crystallising white dwarfs in Sirius-like systems would allow for statistical constraints on crystallisation timescales for the ensemble of white dwarfs.

\section{Conclusions} \label{sec:conclusions}

In this work we have presented the discovery a new Sirius-like system in the solar neighbourhood, composed of a wide quaternary white dwarf companion to the known triple system \thisstar{} \citep{Tokovinin20}. This association was identified through analysis of astrometry from \textit{Gaia}~EDR3, and confirmed beyond doubt using archival proper motion data. By fitting the photometry of \thisstarC{} using state-of-the-art atmosphere models we derive an effective temperature of $6600\pm80$~K and a mass of $0.817\pm0.019~M_\odot$ for the white dwarf, a combination which places it firmly in the parameter space predicted to be occupied by white dwarfs undergoing core crystallisation. This establishes \thisstarC{} as the first \boldagain{confirmed} crystallising white dwarf \boldagain{in} a Sirius-like system\boldagain{.}

\boldagain{\thisstarC{} is} the first known \boldnew{field} white dwarf for which the timescale of crystallisation can be empirically constrained, as the model age of the white dwarf can be compared with the age of the main sequence components to determine whether cooling models adequately reproduce the age of the white dwarf. By combining age estimates from a variety of techniques we measure an age of $7.3^{+1.9}_{-1.8}$~Gyr for the \thisstar{} system, which when compared to our white dwarf age of $4.2\pm0.2$~Gyr results in an age anomaly of $+3.1\pm1.9$~Gyr. This difference is not formally significant ($p=0.05,\:1.7\sigma$), but the mild tension between the age estimates is suggestive of an underestimation of the white dwarf age. We find that this is consistent with the hypothesis of \citet{Blouin21} that phase separation of \Ne{} during core crystallisation can cause a significant delay in the cooling of white dwarfs; for \thisstarC{}, we predict that this process would result in a cooling delay of $\approx$1~Gyr compared to conventional cooling models, which is entirely consistent with our empirical age anomaly. Finally, we propose that the discovery of this system at only 32~parsecs suggests that similar Sirius-like systems containing crystallising white dwarfs are likely to be numerous. Future discoveries may therefore allow for stronger tests of white dwarf crystallisation models.

We conclude that the discovery of the \thisstar{} system has opened up a new avenue for understanding crystallising white dwarfs. We hope that the results of this study will encourage further research for the purpose of identifying and characterising new systems containing crystallising white dwarfs, and that future studies will be able to use these systems to directly constrain theoretical models of core crystallisation.

\section*{Acknowledgements}

We acknowledge and pay respect to Australia’s Aboriginal and Torres Strait Islander peoples, who are the traditional custodians of the lands, waterways and skies all across Australia. We thank the anonymous referee for comments which have helped improve this work. We thank Chelsea Huang and George Zhou for their indispensable help with isochrone fitting. SB is a Banting Postdoctoral Fellow and a CITA National Fellow, supported by the Natural Sciences and Engineering Research Council of Canada (NSERC). This research has made use of the SIMBAD database and VizieR catalogue access tool, operated at CDS, Strasbourg, France. This research has made use of NASA's Astrophysics Data System. This work has made use of the Montreal White Dwarf Database \citep{Dufour2017}. This work has made use of data from the European Space Agency (ESA) mission {\it Gaia} (\url{https://www.cosmos.esa.int/gaia}), processed by the {\it Gaia} Data Processing and Analysis Consortium (DPAC, \url{https://www.cosmos.esa.int/web/gaia/dpac/consortium}). Funding for the DPAC has been provided by national institutions, in particular the institutions participating in the {\it Gaia} Multilateral Agreement. This publication makes use of data products from the Wide-field Infrared Survey Explorer, which is a joint project of the University of California, Los Angeles, and the Jet Propulsion Laboratory/California Institute of Technology, funded by the National Aeronautics and Space Administration.

\section*{Data Availability}

All data used in this work has been collated from publicly accessible depositories and are provided in the text, either in the main body or in the appendices.



\bibliographystyle{mnras}
\bibliography{bib} 

\clearpage




\appendix

\section{White dwarf photometry} \label{appendix:WD_photometry}

\begin{table}
	\centering
	\caption{Photometry of \thisstarC{} used in our atmosphere model.}
	\label{table:photometry_C}
	\begin{tabular}{lrr}
		\hline
		Band & Flux (mJy) & Reference \\
		\hline
		$g$ & $0.631\pm0.006$ & \citet{PanSTARRS1} \\
		$i$ & $0.800\pm0.035$ & \citet{PanSTARRS1} \\
		$z$ & $0.806\pm0.010$ & \citet{PanSTARRS1} \\
		$G$ & $0.662\pm0.002$ & \citet{GaiaEDR3} \\
		$G_{\text{BP}}$ & $0.620\pm0.005$ & \citet{GaiaEDR3} \\
		$G_{\text{RP}}$ & $0.757\pm0.005$ & \citet{GaiaEDR3} \\
		\hline
	\end{tabular}
\end{table}

In Table~\ref{table:photometry_C} we list the photometry of \thisstarC{} used in our atmosphere model.

\section{Isochrone model} \label{appendix:isochrones}

In this section we detail the data and priors used for our isochrone fit of \thisstarAB{}. Through a literature search we have assembled spatially unresolved photometry of the system in the Johnson $B$ and $V$, Tycho $B_\text{T}$ and $V_\text{T}$, \textit{Gaia}~EDR3 $G$, $G_{\text{BP}}$, and $G_{\text{RP}}$, 2MASS $J$, $H$, $K_S$, and WISE $W1$-$W4$ bands. We list this photometry in Table~\ref{table:photometry_AB}. Resolved photometry of the system is limited to the contrast between A and B of 2.5~mag in the $I$ band and 3.9~mag in the $V$ band measured using speckle imaging \citep[][hereafter \citetalias{Tokovinin20}]{Tokovinin20}, for which we assume uncertainties of $\pm$0.1~mag. As \thisstar{}~Ab is unresolved in the speckle observations, the measured magnitude of A is a combination of the fluxes of Aa and Ab.

We assign a Gaussian prior on the stellar distance based on the \textit{Gaia}~EDR3 parallax of \thisstarC{} ($30.911\pm0.063$~mas), while we assume a uniform age prior of $0.1-14$~Gyr. As for priors on the system [Fe/H] and the effective temperature of \thisstarAa{}, \citet{Brewer16} measure $\text{[Fe/H]}=-0.25$ and $T_{\text{eff}}=5604$~K while \citet[][hereafter \citetalias{Fuhrmann19}]{Fuhrmann19} estimate similar values of $\text{[Fe/H]}\approx-0.21$, $T_{\text{eff}}=5650$~K. We choose to adopt the medians of these values ($-0.23$, $5630$~K) as priors for our model, assuming reasonable uncertainties of $\pm$0.05, $\pm$50~K.

\citetalias{Fuhrmann19} additionally estimate effective temperatures for \thisstar{}~Ab and B, which however require some reinterpretation. The authors observed a faint component ($\Delta V\approx4.1$~mag) in their spectrum with a radial velocity anomaly of $\Delta v=+11.4$~km~s$^{-1}$ relative to that of \thisstarAa{}, for which they estimate an effective temperature of $\approx3900$~K. They further found that an additional, un-shifted source is required to produce an internally consistent [Fe/H] abundance, for which they estimate $\Delta V\approx3.4$~mag and $T_{\text{eff}}\approx4100$~K. \citetalias{Fuhrmann19} correctly recognised that their three spectroscopic components are the same as those mentioned in \citet{Tokovinin16}, but without further information they were unable to identify which component is which and thus arbitrarily assigned the redshifted spectroscopic component to Ab and the un-shifted one to B. However, \citetalias{Tokovinin20} note that their orbital solution predicts a radial velocity anomaly of $+10.8$~km~s$^{-1}$ for \thisstar{}~B at that epoch, leading to the conclusion that the fainter, redshifted component of \citetalias{Fuhrmann19} is in fact \thisstar{}~B. The contrast of $\Delta V=3.9$~mag between A and B observed by \citetalias{Tokovinin20} supports this hypothesis, as this is closer to the estimated magnitude contrast of the redshifted component of \citetalias{Fuhrmann19} than of the unshifted component. We therefore assign $T_{\text{eff}}$ priors based on the estimates of \citetalias{Fuhrmann19} based on this identification of the components, which large prior uncertainties of $\pm$250~K. However, we choose not to use the spectroscopic $V$-band contrasts of \citetalias{Fuhrmann19} as priors since it is not clear how precise these estimates can be taken to be.

Some additional priors can be derived from the orbital solution of \citetalias{Tokovinin20}. For the orbit of \thisstar{}~B the authors measure $P=7.446\pm0.025$~yr and $a=0.150\pm0.003$~arcsec; using Kepler's third law and given the system parallax of $\varpi=30.911\pm0.063$~mas we calculate a total mass of $M_{\text{tot}}=2.06^{+0.13}_{-0.12}~M_\odot$, a value which we use as a prior on the total mass of \thisstarAB{}. We are hesitant to use mass priors based on the dynamical component masses since \citetalias{Tokovinin20} point out that light blending from the companions may cause attenuation of the radial velocity signal of Aa. However, based on the astrometric "wobble" of \thisstar{}~B the authors measure a mass ratio for the Aa-Ab subsystem of $0.54\pm0.15$. This is relatively imprecise but is more likely to be accurate, so we use this value as a prior on the inner mass ratio.

\begin{table}
	\centering
	\caption{Photometry of \thisstarAB{} used in the isochrone fit.}
	\label{table:photometry_AB}
	\begin{tabular}{lrr}
		\hline
		Band & Magnitude (mag) & Reference \\
		\hline
		$B$ & $8.39\pm0.02$ & \citet{Lasker08} \\
		$V$ & $7.69\pm0.02$ & \citet{Lasker08} \\
		$B_\text{T}$ & $8.545\pm0.017$ & \citet{Tycho2} \\
		$V_\text{T}$ & $7.764\pm0.011$ & \citet{Tycho2} \\
		$G$ & $7.4965\pm0.0028$ & \citet{GaiaEDR3} \\
		$G_{\text{BP}}$ & $7.8640\pm0.0028$ & \citet{GaiaEDR3} \\
		$G_{\text{RP}}$ & $6.9336\pm0.0038$ & \citet{GaiaEDR3} \\
		$J$ & $6.259\pm0.029$ & \citet{2MASS} \\
		$H$ & $5.884\pm0.024$ & \citet{2MASS} \\
		$K_S$ & $5.767\pm0.020$ & \citet{2MASS} \\
		$W1$ & $5.749\pm0.050$ & \citet{WISE} \\
		$W2$ & $5.621\pm0.025$ & \citet{WISE} \\
		$W3$ & $5.735\pm0.015$ & \citet{WISE} \\
		$W4$ & $5.683\pm0.038$ & \citet{WISE} \\
		\hline
	\end{tabular}
\end{table}

\section{Chemical clock sample} \label{appendix:chemical_clocks}

\begin{table*}
\centering
\caption{Age and abundance data for stars selected from the \citet{Brewer16} sample used to calibrate the chemical clocks.}
\label{table:abundances}
\begin{tabular}{lcccclcccc}
\hline
Name & Age (Gyr) & $\sigma_{\text{Age}}$ (Gyr) & [Mg/Fe] & [Al/Fe] & Name & Age (Gyr) & $\sigma_{\text{Age}}$ (Gyr) & [Mg/Fe] & [Al/Fe] \\
\hline
HD 2589  & 5.2 & 0.8   & 0.03 & 0.08 & HD 8375  & 2.3 & 0.25   & 0.02 & 0.08  \\
HD 4307  & 7.7 & 0.7   & 0.00 & -0.03 & HD 15928 & 3.3 & 0.65   & -0.04 & -0.08 \\
HD 8574  & 4.2 & 0.7   & -0.01 & -0.04 & HD 18015 & 2.8 & 0.5   & -0.06 & -0.12 \\
HD 15335 & 5.7 & 0.35   & -0.02 & -0.05 & HD 19019 & 1.2 & 0.9   & -0.02 & -0.14 \\
HD 21019 & 6.4 & 0.6   & 0.01 & 0.02 & HD 19522 & 3.0 & 0.7   & -0.03 & -0.04 \\
HD 24892 & 9.9 & 1.0   & 0.12 & 0.13 & HD 31543 & 1.6 & 0.55   & -0.10 & 0.01  \\
HD 32923 & 9.1 & 0.65   & 0.08 & 0.10 & HD 45210 & 2.3 & 0.4   & -0.09 & -0.17 \\
HD 33021 & 8.6 & 0.65   & 0.10 & 0.13 & HD 46588 & 2.1 & 0.8   & -0.03 & -0.19 \\
HD 34721 & 4.7 & 0.65   & -0.01 & -0.01 & HD 49933 & 1.6 & 0.35   & 0.02 & --   \\
HD 34745 & 3.5 & 0.85   & -0.01 & -0.06 & HD 72440 & 2.7 & 0.8   & -0.10 & -0.15 \\
HD 35974 & 6.0 & 0.5   & -0.02 & -0.09 & HD 82328 & 3.1 & 0.3   & -0.10 & --   \\
HD 38949 & 1.0 & 0.8   & -0.05 & -0.16 & HD 85472 & 3.8 & 0.5   & -0.05 & -0.10 \\
HD 45067 & 5.0 & 0.3   & -0.05 & -0.10 & HD 95622 & 4.4 & 0.95   & -0.02 & -0.08 \\
HD 48938 & 6.4 & 1.0   & 0.04 & -0.09 & HD 102444 & 3.0 & 0.6   & -0.08 & -0.13 \\
HD 50806 & 9.1 & 0.6   & 0.08 & 0.15 & HD 103616 & 3.4 & 0.9   & -0.03 & 0.12  \\
HD 50639 & 2.7 & 0.65   & -0.02 & -0.10 & HD 103890 & 3.0 & 0.5   & -0.05 & -0.11 \\
HD 67767 & 2.5 & 0.2   & -0.04 & -0.05 & HD 104389 & 1.2 & 1.0   & -0.03 & -0.07 \\
HD 69897 & 3.2 & 0.5   & 0.02 & -0.19 & HD 104860 & 1.4 & 1.0   & -0.10 & -0.22 \\
HD 84117 & 2.6 & 0.55   & -0.02 & -0.08 & HD 108189 & 1.4 & 0.3   & -0.12 & -0.17 \\
HD 95128 & 5.2 & 1.0   & 0.00 & 0.01 & HD 109159 & 3.6 & 0.7   & -0.04 & -0.06 \\
HD 101472 & 0.7 & 0.55   & -0.05 & -0.14 & HD 109218 & 4.4 & 0.8   & -0.03 & -0.07 \\
HD 117176 & 7.8 & 0.55   & 0.01 & 0.01 & HD 120064 & 1.2 & 0.1   & -0.15 & --   \\
HD 141004 & 4.8 & 0.65   & -0.01 & 0.03 & HD 150706 & 1.1 & 0.85   & -0.07 & -0.13 \\
HD 159868 & 5.7 & 0.7   & -0.02 & -0.01 & HD 168151 & 2.7 & 0.55   & -0.01 & --   \\
HD 167665 & 3.9 & 0.7   & 0.00 & -0.08 & HD 171264 & 2.7 & 0.5   & -0.10 & -0.11 \\
HD 168443 & 9.8 & 0.7   & 0.05 & 0.12 & HD 182189 & 9.6 & 0.9   & 0.10 & 0.21  \\
HD 179957 & 7.9 & 0.55   & 0.03 & 0.08 & HD 182736 & 4.3 & 0.45   & -0.05 & -0.06 \\
HD 187923 & 8.8 & 0.65   & 0.07 & 0.13 & HD 183473 & 3.0 & 0.35   & -0.05 & -0.08 \\
HD 188512 & 4.4 & 0.75   & -0.03 & -0.04 & HD 231701 & 3.4 & 0.7   & -0.04 & -0.07 \\
HD 190228 & 5.9 & 0.9   & 0.01 & 0.00 & HD 187013 & 3.7 & 0.3   & -0.06 & -0.15 \\
HD 195564 & 7.4 & 0.45   & -0.02 & -0.03 & HD 187637 & 2.6 & 0.6   & -0.03 & -0.13 \\
HD 198802 & 4.7 & 0.3   & -0.05 & -0.05 & HD 199260 & 1.4 & 0.7   & -0.09 & --   \\
HD 209253 & 0.8 & 0.55   & -0.11 & --  & HD 203471 & 2.4 & 0.65   & -0.06 & -0.10 \\
HD 209458 & 2.5 & 0.95   & -0.01 & -0.07 & HD 210027 & 1.4 & 0.4   & 0.01 & -0.25 \\
HD 236427 & 3.8 & 0.75   & -0.06 & -0.07 & HD 215049 & 3.7 & 0.85   & -0.06 & -0.05 \\
HD 2946  & 2.4 & 0.55   & -0.04 & -0.14 & HD 220554 & 4.3 & 0.85   & 0.14 & 0.23  \\
HD 4395  & 3.2 & 0.45   & -0.02 & -0.12 \\
\hline
\end{tabular}
\end{table*}

Here we describe our sample selection for our chemical clock sample. We take the stellar data from \citet{Brewer16} and apply the following cuts to their sample:

\begin{itemize}
    \item $\text{Age}<10$~Gyr
    \item $\sigma_{\text{Age}}\leq1$~Gyr
    \item $\text{[Fe/H]}<0.05$
    \item $\log g>3.5$~cm~s$^{-2}$
    \item $\text{C-rms}<0.03$
    \item $\text{L-rms}<0.03$.
\end{itemize}

We remove stars with isochronal ages above 10~Gyr from our sample in an attempt to avoid thick disk stars, as these are known to follow age-abundance relations differing from those of thin disk stars \citep{Spina18, Titarenko19}. \citet{Brewer16} report separate positive and negative bounds on their stellar ages; to produce a single value for the age uncertainty we take the median of the range of these values. We select for stars with $\text{[Fe/H]}<0.05$ in recognition of the metallicity dependence of chemical clocks \citep{Feltzing17, DelgadoMena19}, as \thisstarAa{} is a metal-poor star ($\text{[M/H]}=-0.24$ in \citealt{Brewer16}); the specific cutoff was chosen arbitrarily as a compromise between sample size and similarity in metallicity since the number of metal-poor stars in the \citet{Brewer16} sample is relatively small. We remove stars with $\log g<3.5$ to avoid any potential systematic influence of surface gravity on the abundance measurements relative to \thisstarAa{}, which is a dwarf star. Finally, the continuum-RMS and line-RMS values reported by \citet{Brewer16} broadly reflect the precision of the spectroscopic fit, so we select for stars with low values in these parameters.

The sample of 73 stars which survive these cuts are listed in Table~\ref{table:abundances}, along with their ages and abundance ratios from \citet{Brewer16}.


\bsp	
\label{lastpage}
\end{document}